\documentclass[reprint,amsmath,amssymb,aps,prd,superscriptaddress,longbibliography,]{revtex4-2}

\usepackage{graphicx}

\usepackage{dcolumn}
\usepackage{bm}
\usepackage{comment}
\usepackage{xspace}
\usepackage{units}
\usepackage{float}
\usepackage{multirow}
\usepackage{xcolor}
\usepackage[caption=false]{subfig}
\usepackage{amssymb}
\usepackage{enumitem}
\usepackage{amsmath}
\usepackage{commath}
\usepackage{verbatim}
\usepackage{afterpage}
\usepackage{soul}
\usepackage[normalem]{ulem}

\usepackage[colorlinks = true,
            linkcolor = red,
            urlcolor  = blue,
            citecolor = red,
            anchorcolor = blue]{hyperref}

\newcommand{\ANL}{Argonne National Laboratory (ANL), Lemont, IL, 60439, USA}
\newcommand{\Bern}{Universit{\"a}t Bern, Bern CH-3012, Switzerland}
\newcommand{\BNL}{Brookhaven National Laboratory (BNL), Upton, NY, 11973, USA}
\newcommand{\UCSB}{University of California, Santa Barbara, CA, 93106, USA}
\newcommand{\Cambridge}{University of Cambridge, Cambridge CB3 0HE, United Kingdom}
\newcommand{\CIEMAT}{Centro de Investigaciones Energ\'{e}ticas, Medioambientales y Tecnol\'{o}gicas (CIEMAT), Madrid E-28040, Spain}
\newcommand{\Chicago}{University of Chicago, Chicago, IL, 60637, USA}
\newcommand{\Cincinnati}{University of Cincinnati, Cincinnati, OH, 45221, USA}
\newcommand{\CSU}{Colorado State University, Fort Collins, CO, 80523, USA}
\newcommand{\Columbia}{Columbia University, New York, NY, 10027, USA}
\newcommand{\Edinburgh}{University of Edinburgh, Edinburgh EH9 3FD, United Kingdom}
\newcommand{\FNAL}{Fermi National Accelerator Laboratory (FNAL), Batavia, IL 60510, USA}
\newcommand{\Granada}{Universidad de Granada, Granada E-18071, Spain}
\newcommand{\Harvard}{Harvard University, Cambridge, MA 02138, USA}
\newcommand{\IIT}{Illinois Institute of Technology (IIT), Chicago, IL 60616, USA}
\newcommand{\KSU}{Kansas State University (KSU), Manhattan, KS, 66506, USA}
\newcommand{\Lancaster}{Lancaster University, Lancaster LA1 4YW, United Kingdom}
\newcommand{\LANL}{Los Alamos National Laboratory (LANL), Los Alamos, NM, 87545, USA}
\newcommand{\Louisiana}{Louisiana State University, Baton Rouge, LA, 70803, USA}
\newcommand{\Manchester}{The University of Manchester, Manchester M13 9PL, United Kingdom}
\newcommand{\MIT}{Massachusetts Institute of Technology (MIT), Cambridge, MA, 02139, USA}
\newcommand{\Michigan}{University of Michigan, Ann Arbor, MI, 48109, USA}
\newcommand{\Minnesota}{University of Minnesota, Minneapolis, MN, 55455, USA}
\newcommand{\NMSU}{New Mexico State University (NMSU), Las Cruces, NM, 88003, USA}
\newcommand{\Oxford}{University of Oxford, Oxford OX1 3RH, United Kingdom}
\newcommand{\Pitt}{University of Pittsburgh, Pittsburgh, PA, 15260, USA}
\newcommand{\Rutgers}{Rutgers University, Piscataway, NJ, 08854, USA}
\newcommand{\SLAC}{SLAC National Accelerator Laboratory, Menlo Park, CA, 94025, USA}
\newcommand{\SDSMT}{South Dakota School of Mines and Technology (SDSMT), Rapid City, SD, 57701, USA}
\newcommand{\Maine}{University of Southern Maine, Portland, ME, 04104, USA}
\newcommand{\Syracuse}{Syracuse University, Syracuse, NY, 13244, USA}
\newcommand{\TelAviv}{Tel Aviv University, Tel Aviv, Israel, 69978}
\newcommand{\Tennessee}{University of Tennessee, Knoxville, TN, 37996, USA}
\newcommand{\UTA}{University of Texas, Arlington, TX, 76019, USA}
\newcommand{\Tufts}{Tufts University, Medford, MA, 02155, USA}
\newcommand{\UCL}{University College London, London WC1E 6BT, United Kingdom}
\newcommand{\VTech}{Center for Neutrino Physics, Virginia Tech, Blacksburg, VA, 24061, USA}
\newcommand{\Warwick}{University of Warwick, Coventry CV4 7AL, United Kingdom}
\newcommand{\Yale}{Wright Laboratory, Department of Physics, Yale University, New Haven, CT, 06520, USA}

\begin{document}

\title{First demonstration of $\mathcal{O}$(\unit[1]{ns}) timing resolution in the MicroBooNE liquid argon time projection chamber}

\affiliation{\ANL}
\affiliation{\Bern}
\affiliation{\BNL}
\affiliation{\UCSB}
\affiliation{\Cambridge}
\affiliation{\CIEMAT}
\affiliation{\Chicago}
\affiliation{\Cincinnati}
\affiliation{\CSU}
\affiliation{\Columbia}
\affiliation{\Edinburgh}
\affiliation{\FNAL}
\affiliation{\Granada}
\affiliation{\Harvard}
\affiliation{\IIT}
\affiliation{\KSU}
\affiliation{\Lancaster}
\affiliation{\LANL}
\affiliation{\Louisiana}
\affiliation{\Manchester}
\affiliation{\MIT}
\affiliation{\Michigan}
\affiliation{\Minnesota}
\affiliation{\NMSU}
\affiliation{\Oxford}
\affiliation{\Pitt}
\affiliation{\Rutgers}
\affiliation{\SLAC}
\affiliation{\SDSMT}
\affiliation{\Maine}
\affiliation{\Syracuse}
\affiliation{\TelAviv}
\affiliation{\Tennessee}
\affiliation{\UTA}
\affiliation{\Tufts}
\affiliation{\UCL}
\affiliation{\VTech}
\affiliation{\Warwick}
\affiliation{\Yale}

\author{P.~Abratenko} \affiliation{\Tufts}
\author{O.~Alterkait} \affiliation{\Tufts}
\author{D.~Andrade~Aldana} \affiliation{\IIT}
\author{J.~Anthony} \affiliation{\Cambridge}
\author{L.~Arellano} \affiliation{\Manchester}
\author{J.~Asaadi} \affiliation{\UTA}
\author{A.~Ashkenazi}\affiliation{\TelAviv}
\author{S.~Balasubramanian}\affiliation{\FNAL}
\author{B.~Baller} \affiliation{\FNAL}
\author{G.~Barr} \affiliation{\Oxford}
% ...

\collaboration{The MicroBooNE Collaboration}
\thanks{microboone\_info@fnal.gov}\noaffiliation

\date{\today}

\begin{abstract}
MicroBooNE is a neutrino experiment located in the Booster Neutrino Beamline (BNB) at Fermilab, which collected data from 2015 to 2021. MicroBooNE's liquid argon time projection chamber (LArTPC) is accompanied by a photon detection system consisting of 32 photomultiplier tubes used to measure the argon scintillation light and determine the timing of neutrino interactions. Analysis techniques combining light signals and reconstructed tracks are applied to achieve a neutrino interaction time resolution of $\mathcal{O}$(\unit[1]{ns}). The result obtained allows MicroBooNE to access the nanosecond beam structure of the BNB for the first time. The timing resolution achieved will enable significant enhancement of cosmic background rejection for all neutrino analyses. Furthermore, the ns timing resolution opens new avenues to search for long-lived-particles such as heavy neutral leptons in MicroBooNE, as well as in future large LArTPC experiments, namely the SBN program and DUNE.
\end{abstract}

\setlength{\abovecaptionskip}{3pt}
\setlength{\belowcaptionskip}{3pt}

\addtolength{\tabcolsep}{5pt}
\maketitle

\section{Introduction}
\label{sec:introduction}

The Standard Model (SM) of particle physics has demonstrated remarkable success in describing the interactions between observed fundamental particles; yet clear gaps remain in our ability to address questions such as the nature of dark matter or the matter-antimatter asymmetry in our universe. The study of neutrino properties and oscillations provides a compelling avenue both to complete our understanding of the SM and to explore physics Beyond the Standard Model (BSM). An extensive experimental program comprised of the Deep Underground Neutrino Experiment (DUNE)~\cite{bib:DUNE} and Short Baseline Neutrino (SBN) program~\cite{bib:SBN} intends to make precision measurements of neutrino oscillations using liquid argon time projection chambers (LArTPCs). These detectors offer the ideal environment in which to search for BSM physics in the sub-GeV energy regime. Yet, fully exploiting the potential of such detectors for BSM searches requires dedicated advances in analysis tools and techniques. While millimeter-level accuracy and detailed calorimetric information have enabled the delivery of precision neutrino physics measurements with TPCs~\cite{bib:uBLEE,bib:PeLEE,bib:WCLEE,bib:DLLEE,bib:gLEE,bib:ArCC}, the use of scintillation light signals has not yet been exploited as extensively.
\par This paper presents the first demonstration of $\mathcal{O}$(\unit[1]{ns}) timing resolution for neutrino interactions in a LArTPC utilizing the MicroBooNE detector. This work significantly improves on MicroBooNE's previously reported~\cite{bib:NUID} timing resolution of $\mathcal{O}$(\unit[100]{ns}). A correction to the reconstructed interaction time is applied by introducing four developments: incorporating more precise beam timing signals from the accelerator, improving the reconstruction of signals from MicroBooNE's photon detection system, considering the particle and light propagation in the detector, and, finally, including an empirical calibration to correct for non-uniformities in detector response and particle propagation time.
\par The significance of this analysis has strong implications for searches for BSM physics that exploit differences in time-of-flight (ToF) to detect massive long-lived particles arriving at the detector delayed with respect to neutrinos. The techniques described in this article will allow improved searches beyond those already achieved with MicroBooNE previous analysis~\cite{bib:A2,HNL1}. Furthermore, improved timing can add a new tool for cosmic background rejection in surface LArTPCs, orthogonal to existing techniques~\cite{bib:SBN,bib:ArCC,bib:FM1,bib:FM2}.
\par The remainder of this paper is arranged as follows: Section~\ref{sec:uBintro} provides an overall description of the MicroBooNE detector and the Booster Neutrino Beamline (BNB). Section~\ref{sec:beamrec} describes the analysis developed to demonstrate MicroBooNE's $\mathcal{O}$(\unit[1]{ns}) timing resolution. Section~\ref{sec:results} summarizes the analysis results. Section~\ref{sec:app} presents two applications in which the timing resolution achieved can improve MicroBooNE’s capability of studying neutrino interactions: introducing a new tool for cosmic background rejection and improving the performance for BSM physics searches.

\begin{figure*}
\centering
\includegraphics[width=0.65\textwidth]{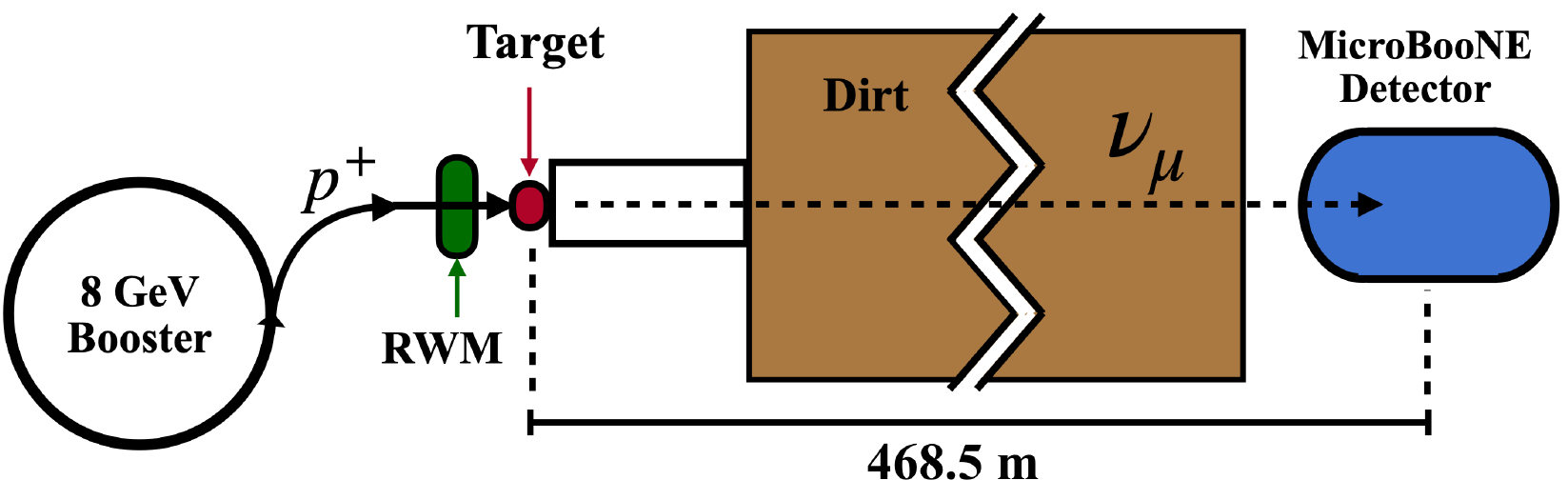}
\caption{\label{fig:fullbeamline}Schematic of the BNB and MicroBooNE detector. MicroBooNE's detector is in the path of the BNB, on axis with the beam direction, \unit[468.5]{m} downstream of the proton target (red). The RWM (green) records the proton pulse shape immediately before protons hit the target. For events selected in this analysis, the time for protons to hit the target, the propagation and decay of mesons, and the travel time of neutrinos to the detector upstream wall is assumed the same for each event.}
\end{figure*}
\begin{figure*}
\centering
\includegraphics[width=0.95\textwidth]{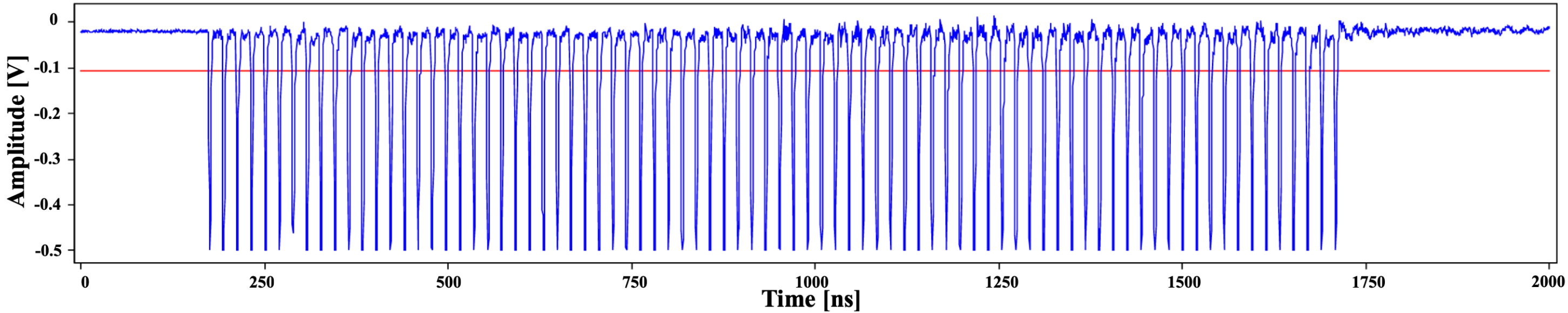}
\caption{\label{fig:RWMstructure}Trace of a single BNB RWM waveform showing the BNB ns substructure. The red line shows the discriminator threshold used by the oscilloscope. The waveform sample frequency is 2 GHz. The vertical axis is the induced charge on the RWM in volts. Each BNB proton pulse is composed of 81 bunches spaced at \unit[$\Delta$~=~18.936~$\pm$~0.001]{ns}. The average bunch width is \unit[$\langle \sigma_{BNB} \rangle$~=~1.308~$\pm$~0.001]{ns}. The RWM time structure shown in this figure is obtained through the instruments and methods described in~\cite{BNB4}.}
\end{figure*}

\vspace{-1em}
\section{Booster Neutrino Beamline and MicroBooNE detector}
\label{sec:uBintro}

MicroBooNE~\cite{uBDesign} is a neutrino experiment at Fermilab that collected data from 2015 to 2021. The detector consists of a LArTPC located near the surface, on axis with the neutrino beam, and \unit[468.5]{m} downstream of the proton target. Figure~\ref{fig:fullbeamline} shows a schematic of the BNB and MicroBooNE detector, which will be briefly described in this section.
\par \emph{Booster Neutrino Beamline.} The primary source of neutrinos for the MicroBooNE experiment is the neutrino beam produced by the BNB~\cite{BNB1}, where \unit[8]{GeV} (kinetic energy) proton pulses are extracted from the Booster accelerator and delivered to the target. Each proton pulse has a \unit[52.81]{MHz} substructure with 81 bunches spaced at \unit[$18.936\pm 0.001\,$]{ns}. The average bunch width is \unit[$\langle \sigma_{BNB} \rangle = 1.308 \pm 0.001\,$]{ns}~\cite{BNB4}. This characteristic sub-structure is key to leveraging ns-scale timing resolution for neutrino interactions, as it leads to wide gaps between neutrino bunches~\cite{BNB2}. 
\par
\begin{figure}[H]
\vspace{-1em}
\centering
\includegraphics[width=0.48\textwidth]{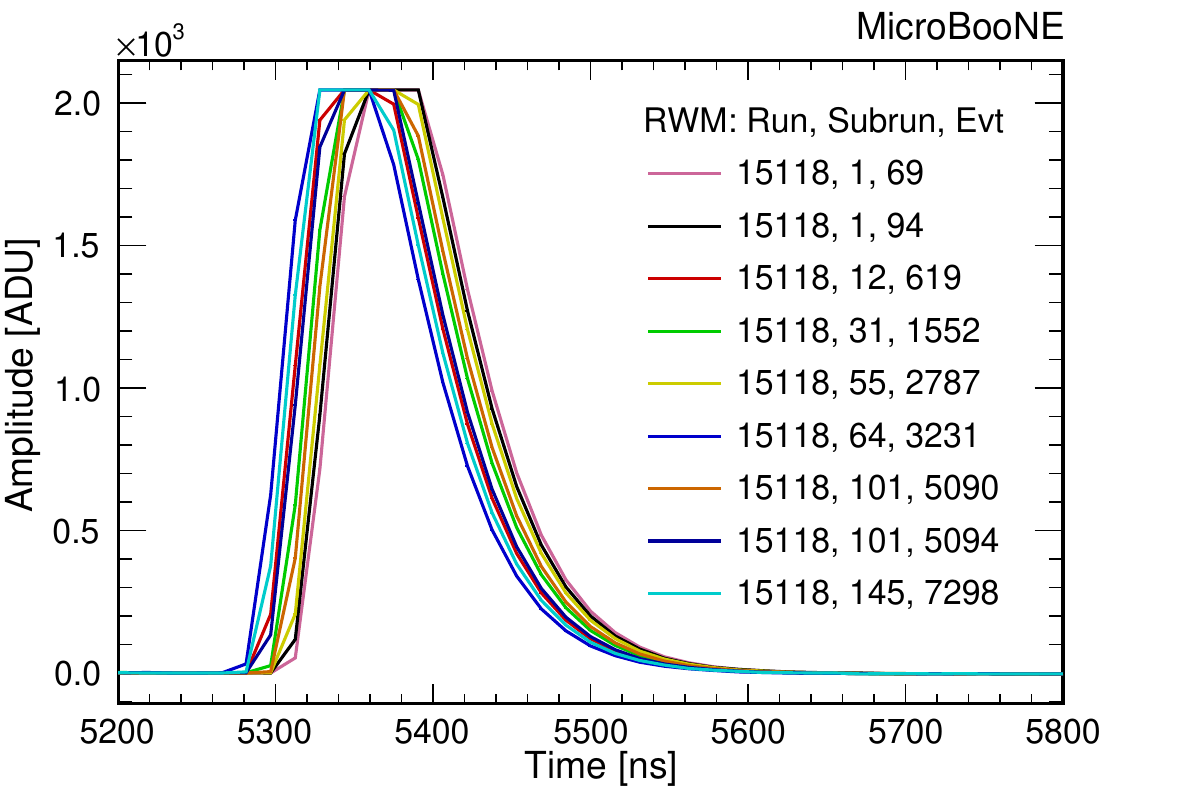}
\caption{\label{fig:RWMpulse}RWM logic pulses in coincidence with the first proton bunch from the accelerator as recorded by the MicroBooNE DAQ. Misalignment between the pulses reflects the main trigger jitter.}
\end{figure}

 \emph{Resistive wall current monitor.} The BNB trigger in MicroBooNE is provided by a copy of the signal coordinating the proton pulse extraction from the Booster accelerator. That signal is subject to a relatively large jitter, which has a fluctuation of tens of ns. To improve on the timing accuracy of the MicroBooNE beam trigger this analysis makes use of the resistive wall current monitor (RWM)~\cite{BNB4} signal. Charged particles traveling through a conductive metallic pipe induce an image current on the pipe wall. In the BNB, the RWM is located just before the proton target and measures the image current produced by the beam protons. The RWM current reproduces accurately the proton pulse's longitudinal time profile. A typical waveform from the BNB RWM, digitized at \unit[2]{GHz}, is shown in Fig.~\ref{fig:RWMstructure}. The first bunch of this signal is used to send a thresholded logic pulse to the MicroBooNE readout electronics where it is recorded for offline monitoring. Figure~\ref{fig:RWMpulse} shows examples of RWM logic pulses recorded with MicroBooNE’s electronics. Misalignment between the pulses reflects the jitter of the BNB trigger. 
\vspace{-1em}
\par
\begin{figure}[H]
\vspace{-1em}
\centering
\includegraphics[width=0.48\textwidth]{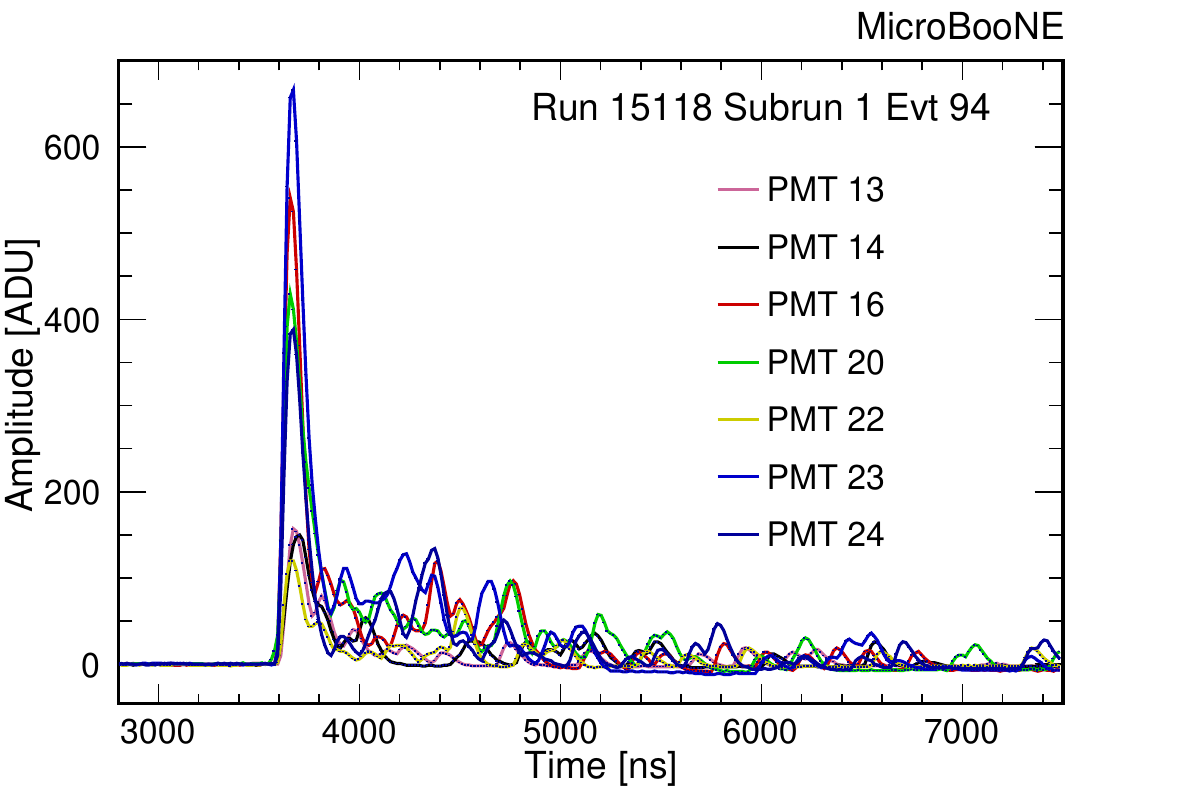}
\caption{\label{fig:pmtwf1}PMT waveforms for a typical neutrino candidate. A subset of the 31 waveforms recorded and a reduced time window around the event is shown.}
\end{figure}

 \emph{MicroBooNE's photon detection system.} A photon detection system~\cite{pmteff} is installed behind the TPC anode plane to detect scintillation light emitted by the argon atoms that are excited by charged particles passing through the argon. Liquid argon is a high-performance prompt scintillator with a yield of about \unit[30,000]{photons/MeV} at MicroBooNE's nominal electric field of \unit[273]{V/cm}~\cite{franciole,DOKE} with $\sim\!\!23\%$ of the total light emitted within a few ns~\cite{Acciarri_2010}. The MicroBooNE photon detection system consists of 32 8-inch cryogenic Hamamatsu photomultiplier tubes (PMTs) equipped with wavelength-shifting tetraphenyl butadiene (TPB) coated acrylic front plates~\cite{pmteff}. MicroBooNE's readout electronics~\cite{Kaleko} record \unit[23.4]{$\mu$s} long waveforms starting at the beam trigger. PMT pulses are smoothed by an analog unipolar shaper with a \unit[60]{ns} rise time and then digitized at \unit[64]{MHz} (\unit[16.625]{ns} samples). One of the 32 PMT channels became unresponsive starting in the summer of 2017. Figure~\ref{fig:pmtwf1} shows example PMT waveforms of scintillation light produced by a candidate neutrino interaction recorded with the MicroBooNE photon detection system. 


\vspace{-2em}
\section{Data analysis}
\label{sec:beamrec}
\vspace{-1em}
The $\mathcal{O}$(\unit[1]{ns}) timing resolution in MicroBooNE is achieved through four analysis steps. First, the RWM logic pulse is used to remove the BNB trigger jitter. Second, an accurate pulse-fitting method is implemented to extract the arrival time of the first photons detected by MicroBooNE's PMTs. Third, the propagation times of particles and scintillation photons inside the detector are extracted by leveraging the TPC's 3D reconstruction. Finally, an empirical calibration is used to apply corrections on the daughter particles' and scintillation light's propagation times. The dataset used in this analysis is an inclusive selection of $\nu_{\mu}$CC interactions candidates~\cite{numucc} from MicroBooNE's BNB collected in 2016--17. Events are reconstructed with the Pandora multi-purpose pattern-recognition toolkit~\cite{Marshall_2017}. This selection yields an $\mathcal{O}$(80$\%$) pure sample of neutrino interactions, and $\mathcal{O}$(20$\%$) cosmic-ray background. The MicroBooNE timing resolution is evaluated by comparing the reconstructed BNB ns substructure with the waveform provided by the RWM, shown in Fig.~\ref{fig:RWMstructure}. The timing resolution achieved by this analysis resolves for the first time in MicroBooNE the substructure of the BNB beam spill~\cite{BNB2}. This section will describe in detail the analysis steps developed to achieve this result.
\begin{figure}[b]
\vspace{-2em}
\centering
\includegraphics[width=0.48\textwidth]{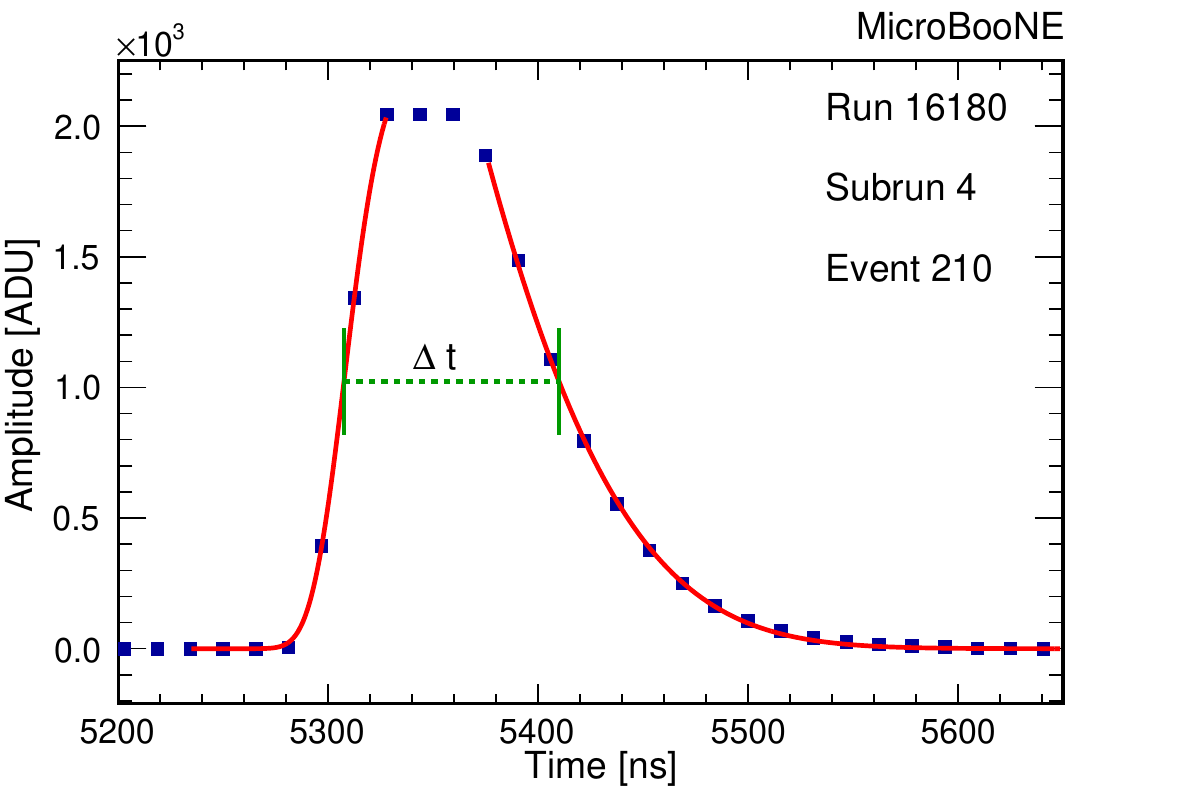}
\centering 
\parbox{0.45\textwidth}{\centering (a) The RWM pulse width ($\Delta t$), shown with the green dotted line, is the distance between the half-height of the rising and falling edges, shown with red curves.}
\includegraphics[width=0.48\textwidth]{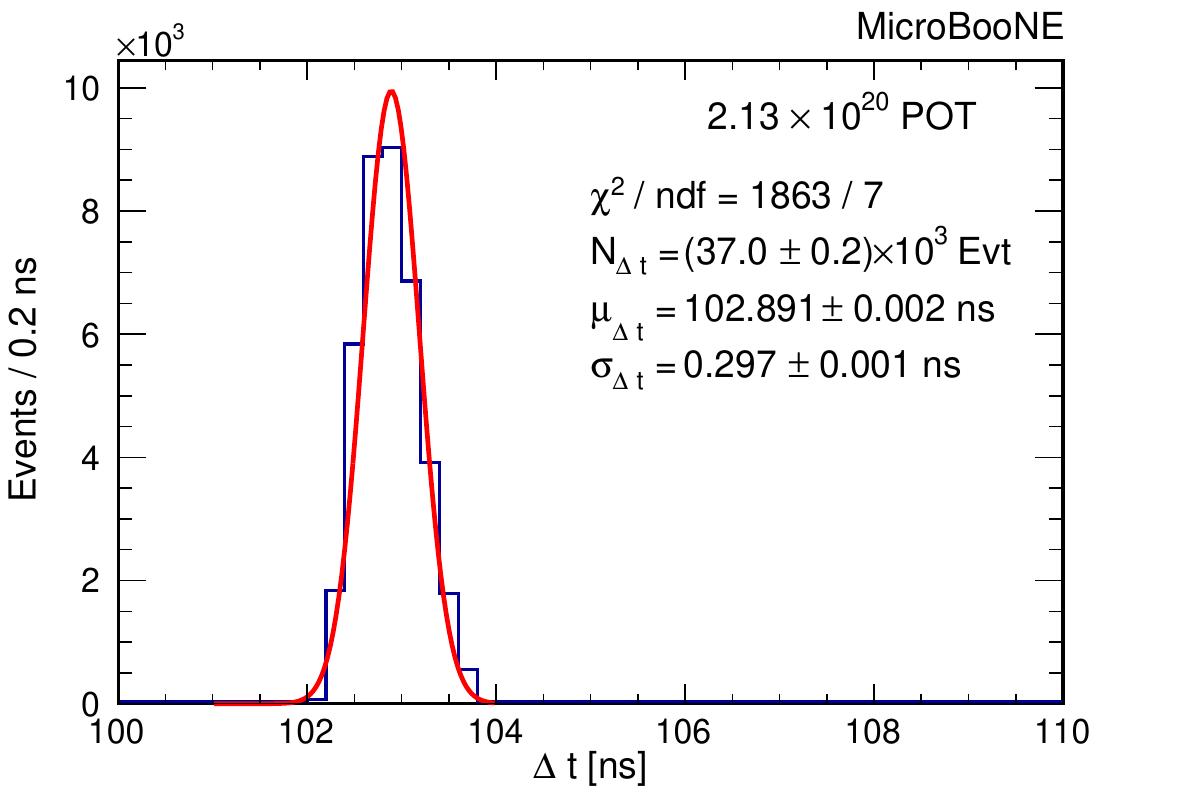}
\centering
\parbox{0.45\textwidth}{\centering (b) Gaussian fit of the $\Delta t$ distribution. The parameters $N_{\Delta t}$, $\mu_{\Delta t}$ and $\sigma_{\Delta t}$ are respectively the normalization, the mean and the standard deviation of the Gaussian fit.}
\caption{The intrinsic timing resolution of the PMT electronics is obtained measuring the stability of the RWM pulse width ($\Delta t$), shown in (a). The $\Delta t$ distribution is fitted with a Gaussian function, shown in (b), and the parameter $\sigma_{\Delta t}/\sqrt{2}$ is used to evaluate the
intrinsic timing resolution of the PMT electronics.}\label{fig:wfres}
\end{figure}
\vspace{1em}
\par \emph{RWM timing.} The RWM logic pulse recorded at MicroBooNE is shaped and digitized through the same readout electronics as the PMTs. The signal timing ($T$\textsubscript{RWM}) is extracted with the fitting method described in the next paragraph. The RWM timing is used to replace the BNB trigger which contains a jitter of tens of ns. The RWM recorded signal is a logic pulse and, therefore, its shape is expected to be stable over time. Because of this, the RWM pulse is used to evaluate the intrinsic timing resolution of the PMT electronics by measuring the stability of the RWM pulse half height width ($\Delta t$), shown in Fig.~\ref{fig:wfres}~(a). The uncertainty of $\Delta t$ is obtained fitting the $\Delta t$ distribution with a Gaussian function, shown in Fig.~\ref{fig:wfres}~(b). The width of the Gaussian ($\sigma_{\Delta t}$) gives the uncertainty of $\Delta t$, which is \unit[$\sigma_{\Delta t}\simeq 0.3$]{ns}. This uncertainty is on the difference between the rising and falling edges of the RWM pulses, both obtained with the same fitting method. Therefore the uncertainty on the single rising edge timing is given by \unit[$\sigma_{\Delta t}$/$\sqrt{2}\simeq 0.2$]{ns}, negligible compared to the overall resolution achieved.

\vspace{1em}
\par
\begin{figure}[H]
\centering
\includegraphics[width=0.48\textwidth]{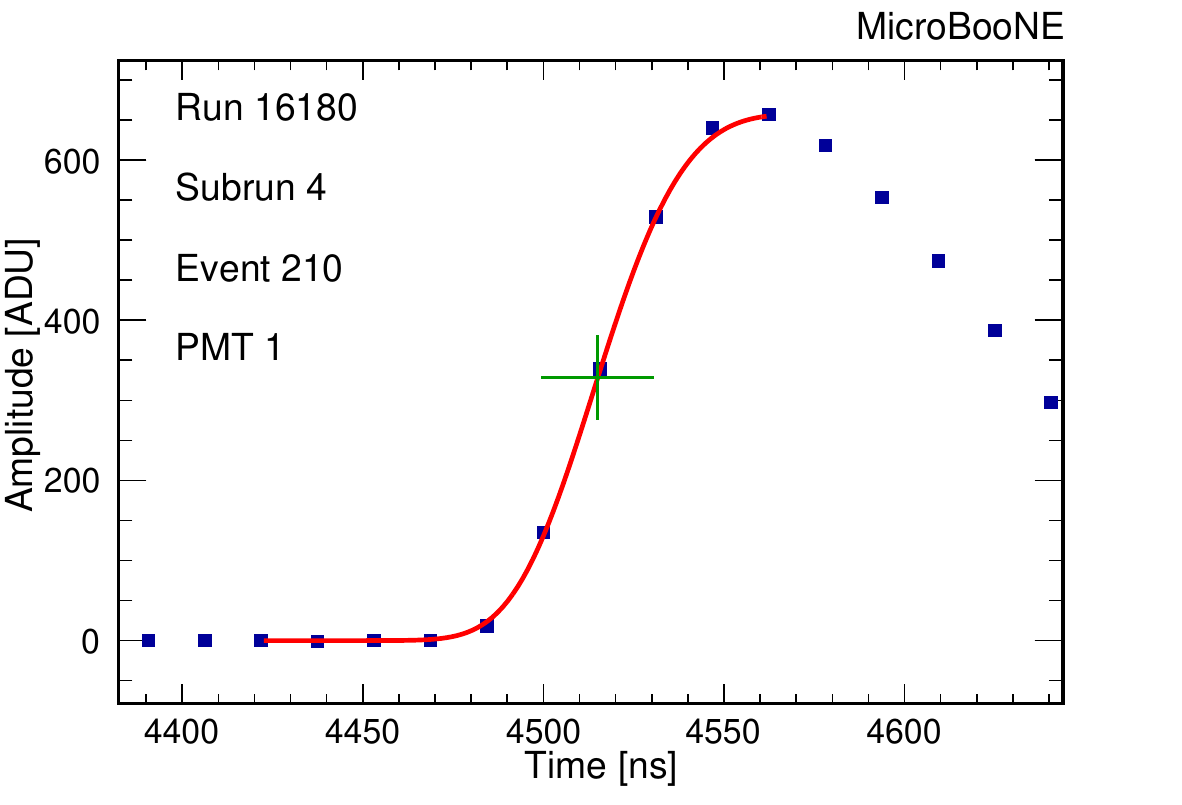}
\caption{\label{fig:pmtwf2}Single PMT pulse timing extraction. The red curve shows the pulse rising-edge fit, and the green cross marks the rising-edge half-height point used to assign the timing to the PMT pulse.}
\end{figure}

 \emph{PMTs Pulse Fitting.} MicroBooNE's PMTs provide a prompt response to the scintillation light produced in neutrino interactions. In order to extract $\mathcal{O}$(\unit[1]{ns}) timing resolution the 60 ns shaping response of the MicroBooNE PMT electronics must be accounted for. This is achieved by fitting the rising edge of the PMT trace with the function 

\begin{equation}
f(t)=A\cdot \exp{\left(- \frac{(t-t_M)^4}{B} \right)}. 
\label{eq:fitFF}
\end{equation} 
\noindent Multiple functions have been tested for fitting the PMT waveform rising edge. The one which gives the lowest $\chi^2$ has been chosen. An example of this fit is shown by the red line in Fig.~\ref{fig:pmtwf2}. The parameters $A$ and $B$ in the fit function are left free and $t_M$ is fixed to the time-tick with the maximum ADC value. The measured half-height value (green cross in Fig.~\ref{fig:pmtwf2}) is used to assign the arrival time of the first photons at the PMT. Despite the relatively low sampling frequency of the PMT digitization, the fitting procedure shows a resolution of \unit[$\simeq\,$0.2]{ns} for the intrinsic timing of the PMT electronics as demonstrated with the RWM pulse.

\begin{figure}[t]
\centering
\includegraphics[width=0.48\textwidth]{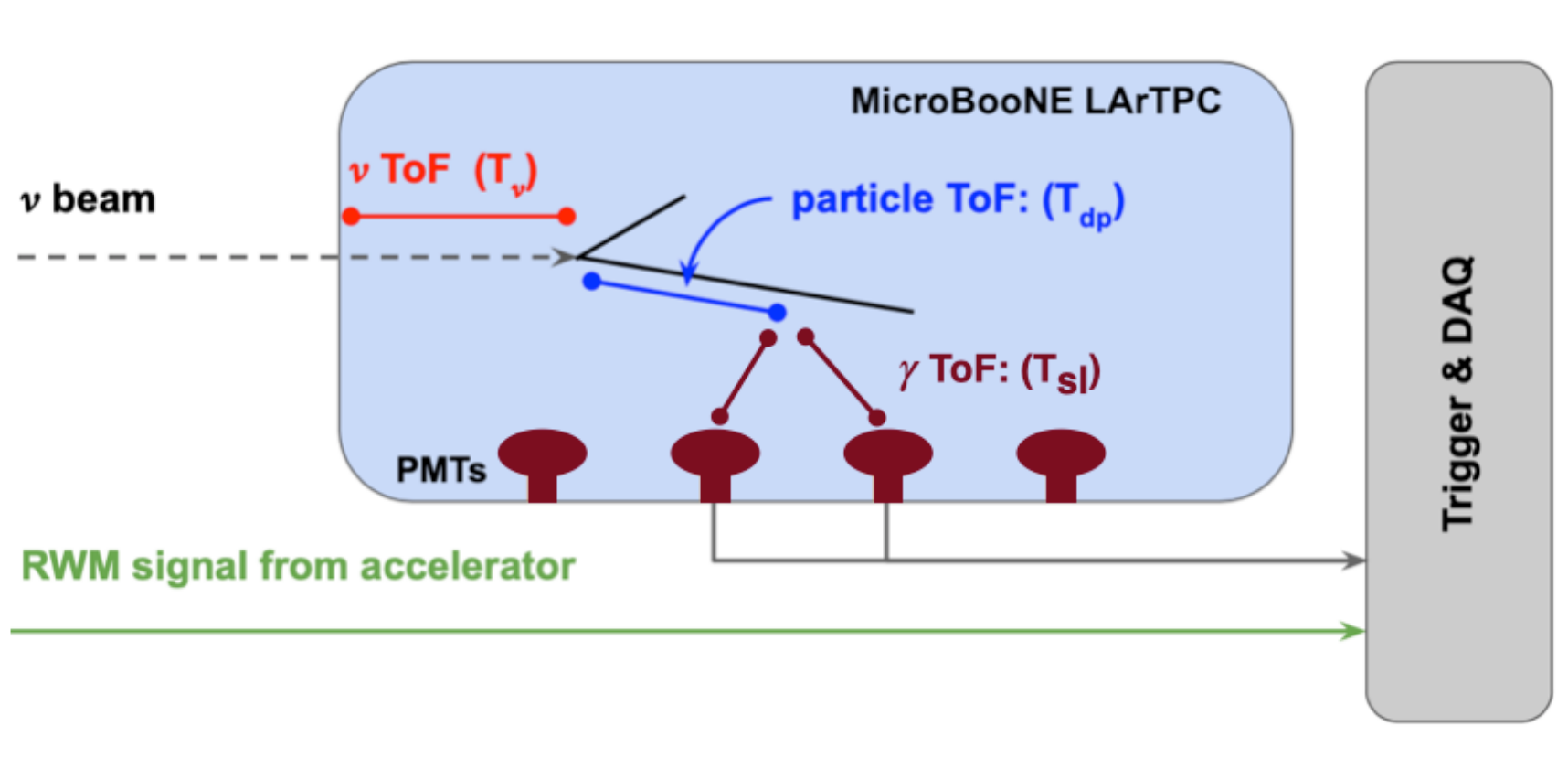}
\caption{\label{fig:uBToF}Schematic of the MicroBooNE LArTPC (light blue). PMTs are represented in maroon. The tracks reconstructed in the TPC (black solid lines) are used to measure the paths of the particles and scintillation photons inside the detector. The three paths, red for the neutrino in the TPC, blue for a daughter particle, and maroon for scintillation photons, are used to evaluate the time between the neutrino entering the TPC and scintillation photons reaching the \vbox{PMTs: $T_{\nu}+T_{dp}+T_{sl}$}.}
\end{figure}
\vspace{-1em}


\par \emph{Particle and scintillation photon propagation.} Between the signal induced by protons at the RWM and the signal provided by PMTs, there is a complex chain of processes to take into account in order to extract the neutrino interaction timing. The time for protons to hit the target, the propagation and decay of mesons, and the travel time of neutrinos to the detector (illustrated in Fig.~\ref{fig:fullbeamline}) is treated as a constant offset for all interactions. Therefore, the neutrino time profile at the upstream detector wall is assumed the same as the proton time profile provided by the RWM. Once neutrinos enter the detector, three processes, shown in Fig.~\ref{fig:uBToF}, impact the observed neutrino interaction time in the PMTs:
\begin{enumerate}
\item The neutrino ToF inside the TPC ($T_{\nu}$);
\item The daughter particle ToF from the neutrino interaction vertex to the space-point where photons are produced ($T_{dp}$); and
\item The scintillation light ToF from the space-point where photons are produced to the PMT where photons are detected ($T_{sl}$).
\end{enumerate}
Leveraging the neutrino interaction vertex position and the daughter particle's track geometry reconstructed with the TPC signals~\cite{Marshall_2017}, the times for each of these three processes can be extracted. Since the beam is on-axis with the detector, and neutrinos are nearly massless, $T_{\nu}$ is given by the neutrino interaction vertex coordinate along the beam direction divided by the speed of light. $T_{dp}$ and $T_{sl}$ are calculated together for all 3D spacepoints along the trajectory of all visible daughter particles from the neutrino interaction. At each 3D spacepoint, $T_{dp}$ is given by the distance from the neutrino interaction vertex divided by the speed of light, and $T_{sl}$ is given by the distance to the TPB coated plate in front of each PMT divided by the group velocity for scintillation light in liquid argon, $v_g$ ($1/v_g=\unit[7.46\pm0.08]{ns/m}$~\cite{Babicz2020}). The minimum value of $T_{dp}+T_{sl}$ among all reconstructed 3D spacepoints is chosen as the daughter particle and scintillation light propagation time for the first photons arriving on the PMT. This quantity is denoted ($T^*_{dp}+T^*_{sl}$). Note that this calculation is performed independently for each PMT. The neutrino ToF inside the TPC ($T_{\nu}$) and the daughter particle and photon propagation times ($T^*_{dp}+T^*_{sl}$) are subtracted from each PMT's measured photon arrival time to obtain the neutrino arrival time at the upstream detector wall. The 81 bunches of the beam pulse sub-structure are visible in the reconstructed neutrino arrival time profile and reproduce the \unit[52.81]{MHz} substructure of the RWM waveform of Fig.~\ref{fig:RWMstructure}.

\vspace{1em}
\par \emph{Empirical calibration.} Once the beam pulse sub-structure can be resolved, measurements of the time distribution of the 81 bunches provide a reference used to empirically correct timing offsets due to non-uniformities in detector response. The 81 bunches are merged in a single peak and a Gaussian fit is performed to extract the mean time $\mu$. Displacements in $\mu$ as a function of a given variable indicate a non-uniformity in need of calibration. Three variables are identified as a source of substantial smearing. 
\begin{enumerate}
\item \emph{PMT hardware.} Variation in signal propagation time due to electronics response, signal transmission, or other intrinsic delays can introduce PMT-by-PMT offsets.
\item \emph{Daughter particle propagation speed.} Approximating the daughter particle velocity to be equal to the speed of light impacts the calculation of the propagation time from the neutrino vertex to each PMT ($T^*_{dp} + T^*_{sl}$). This assumption is adopted because the analysis is implemented prior to detailed particle tracking and identification which would allow to reconstruct the momentum and speed along the trajectory.
\item \emph{Signal amplitude impact on time extraction.} The arrival time is extracted from a fixed amplitude ratio of the waveform rising edge (see Fig.~\ref{fig:pmtwf2}). Although this choice resulted in the best performance, it may introduce a bias dependent on the number of photons collected in the fast component on a given PMT ($N_{ph}$). 
\end{enumerate}

\begin{figure}[h]
\centering
\includegraphics[width=0.48\textwidth]{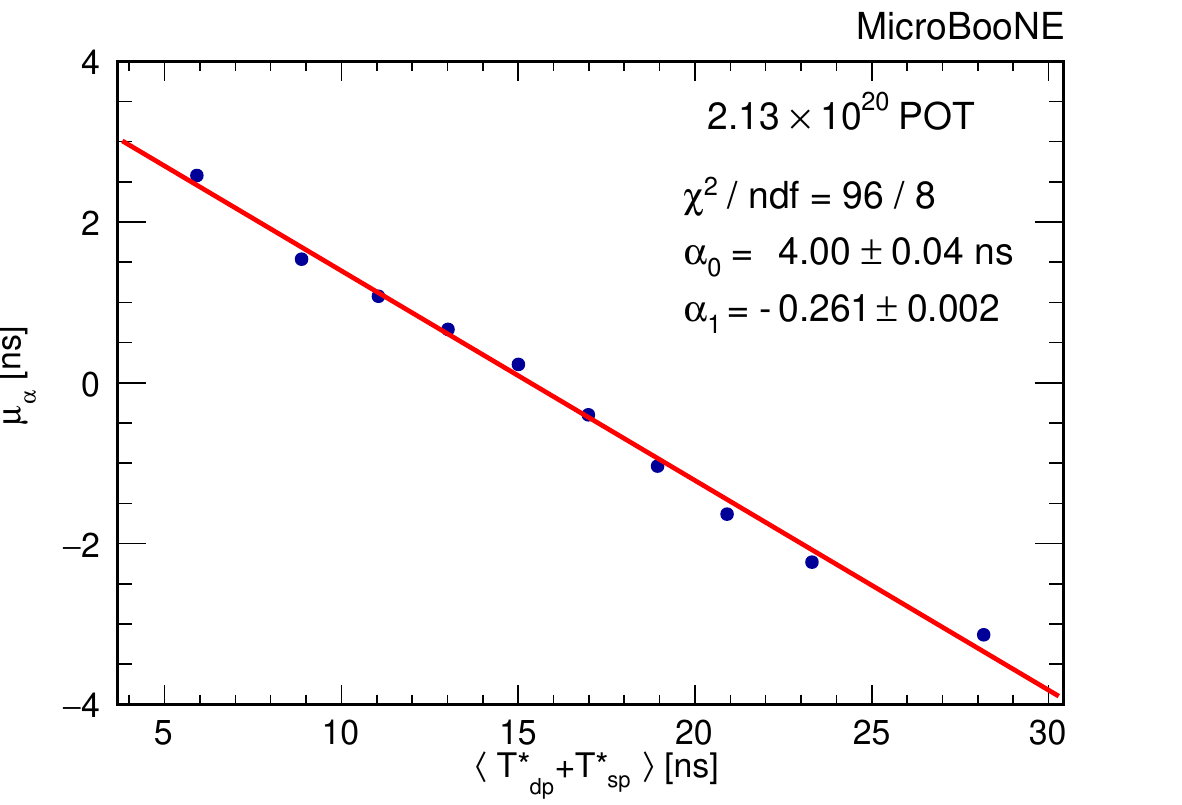}
\centering
\parbox{0.45\textwidth}{\centering(a) Linear fit of the mean of the neutrino interaction time as a function of the average propagation time from the neutrino vertex to a given~PMT~$\langle T^*_{dp}+T^*_{sl}\rangle$. The parameters $\alpha_0$ and $\alpha_1$ are respectively the offset and the gradient.}
\includegraphics[width=0.48\textwidth]{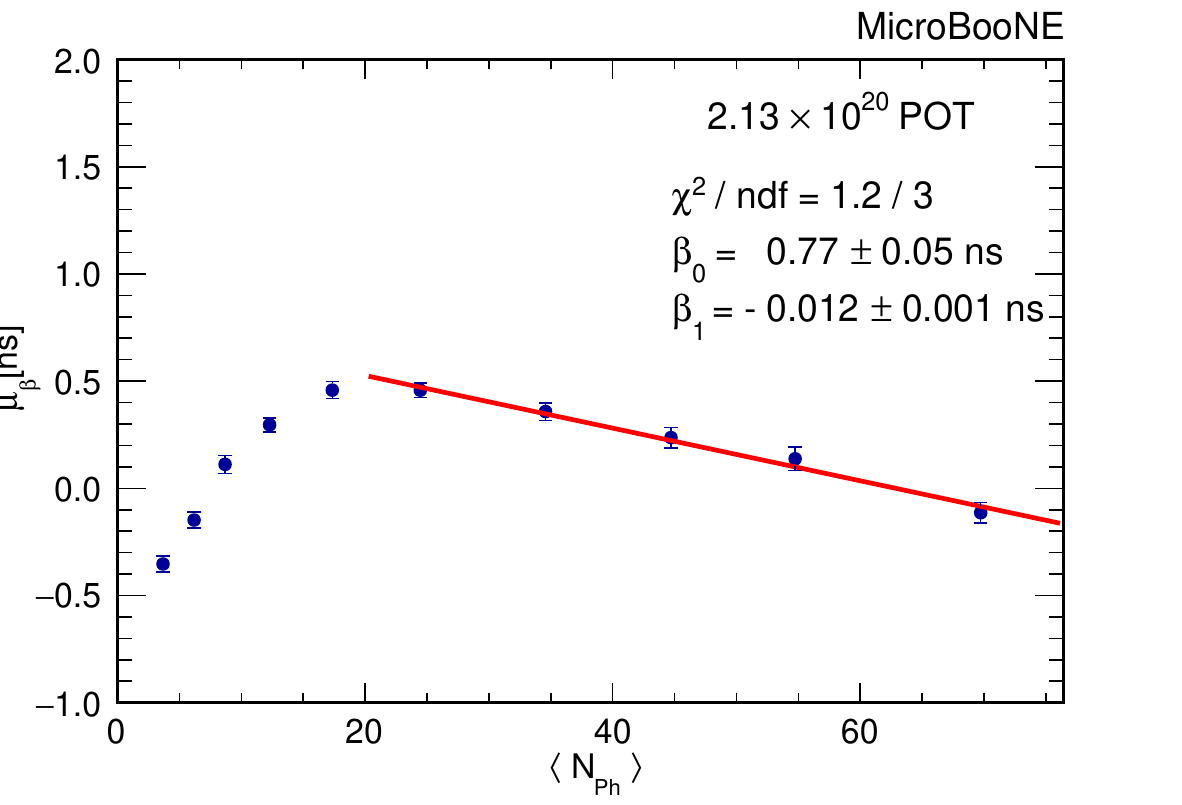}
\centering
\parbox{0.45\textwidth}{\centering(b) Linear fit of the mean neutrino interaction time as a function of the average of number of photons collected by a given~PMT~$\langle N_{Ph} \rangle$. The parameters $\beta_0$ and $\beta_1$ are respectively the offset and the gradient.}
\caption{Linear fits of the mean neutrino interaction time as functions of $\langle T^*_{dp}+T^*_{sl}\rangle$ (a) and $\langle N_{Ph}\rangle$ (b) are used to extract the two calibration factors $\alpha_1$ and $\beta_1$, which are the gradients of the linear fits. The $\beta_1$ coefficient calculation limits the fit to events for which $N_{Ph}$ is larger than 20 photons in order to avoid the introduction of terms above the linear one in the fit function. Nevertheless, the correction is applied to every single PMT measurement.}
\label{fig:ccb}
\end{figure}

\begin{figure*}
\centering
\includegraphics[width=1.0\textwidth]{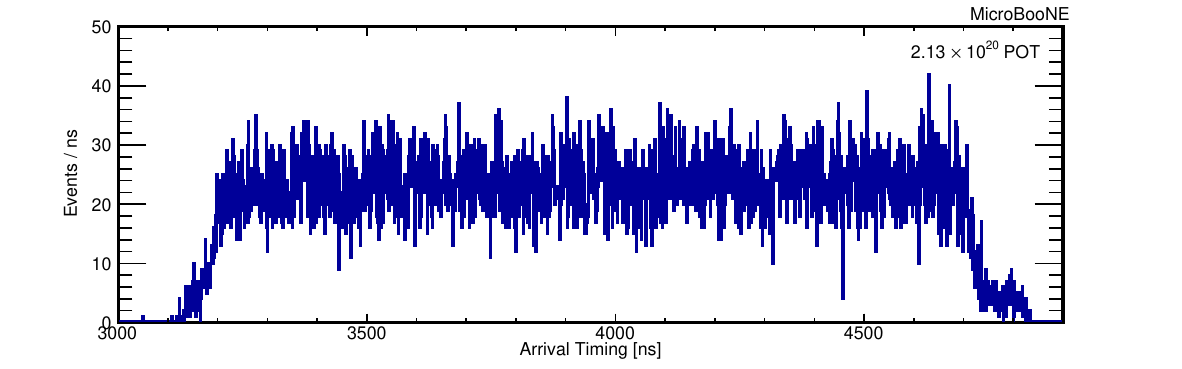} 
\centering
(a)~Neutrino arrival time distribution before the propagation reconstruction.
\includegraphics[width=1.0\textwidth]{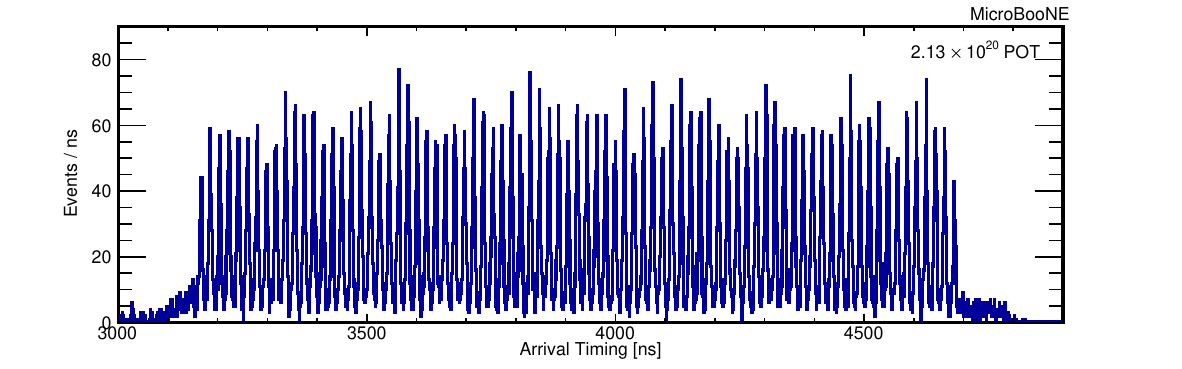}
\centering
(b)~Neutrino arrival time distribution after the propagation reconstruction.
\caption{\label{fig:med0Lb}Neutrino candidate arrival time distribution at the upstream detector wall before (a) and after (b) the propagation reconstruction of the processes happening inside the TPC. The reconstruction includes the neutrino ToF inside the TPC, the daughter particle propagation and the scintillation light propagation, with the relative empirical correction included. The 81 bunches composing the beam pulse sub-structure are easily visible after the propagation reconstruction.}
\end{figure*}

\noindent These three factors are calibrated using the following analysis procedure. First a correction is implemented to account for PMT-by-PMT offsets. The remaining two effects are subsequently calibrated simultaneously. To incorporate a correction for PMT hardware offsets, the value of $\mu$ obtained for each PMT is used to remove the offset with respect to the average across all PMTs. Offsets between PMTs ($T_{os}$) were found to be of order \unit[2.5]{ns}. For the other two factors, the timing distributions are binned once for the propagation time ($T^*_{dp}+T^*_{sl}$) values and once for the number of photons collected in the fast component $N_{Ph}$. Average values $\langle T^*_{dp}+T^*_{sl} \rangle$ and $\langle N_{Ph} \rangle$ and the respective Gaussian means, $\mu_{\alpha}$ and $\mu_{\beta}$, are calculated for each timing distribution. Linear fits of $\mu_{\alpha}$ and $\mu_{\beta}$ as functions of $\langle T^*_{dp}+T^*_{sl} \rangle$ and $\langle N_{Ph} \rangle$ respectively are performed, see Fig.~\ref{fig:ccb}. The fit gradients $\alpha_1$ and $\beta_1$ give the empirical calibration term $T_\mathrm{Emp}=(T^*_{dp}+T^*_{sl}) \cdot \alpha_1 + N_{Ph}\cdot \beta_1$, which is subtracted from the photon arrival time given by each PMT individually. Corrections introduced by the two calibration factors $\alpha_1$ and $\beta_1$ are inversely proportional to each other, causing the spread of the mean values of the beam timing in one variable to increase after a correction for the other variable is applied. For this reason, the corrections are applied simultaneously. The spread as a function of these variables persists after a first correction is applied. To further reduce the residual smearing, the same procedure is repeated. After a few steps, when each iteration is no longer reducing the smearing, the spread of the mean values $\mu_{\alpha}$ and $\mu_{\beta}$, shown in Fig.~\ref{fig:ccb}, is reduced below \unit[0.5]{ns} in both cases.


\begin{table}[b]
\centering
\caption{\label{tab:spreadb}Terms analyzed in the reconstruction steps introduce different contributions to the event timing spread. This table summarizes the standard deviation (STD) and full range of the distribution of values of each term.}
\begin{tabular}{l c r}
 \\
\hline \hline
Term & STD [ns] & Range [ns]\\
 \hline
$T$\textsubscript{RWM}             & $\simeq$ 9   & [-25,+25]  \\
$T$$_{\nu}$             & $\simeq$ 9   & [0, 33]    \\
$ \left( T^*_{dp}+T^*_{sl} \right) $ & $\simeq$ 7   & [0, $>$50]  \\
$T$$_{os}$              &  $\simeq$ 2.5   & [-5, +5]    \\
$T$$_\mathrm{Emp}$            & -   & [-4, +4]    \\
\hline \hline
\end{tabular}
\end{table} 

\vspace{1em}
\par
\begin{figure}[]
\centering
\includegraphics[width=0.48\textwidth]{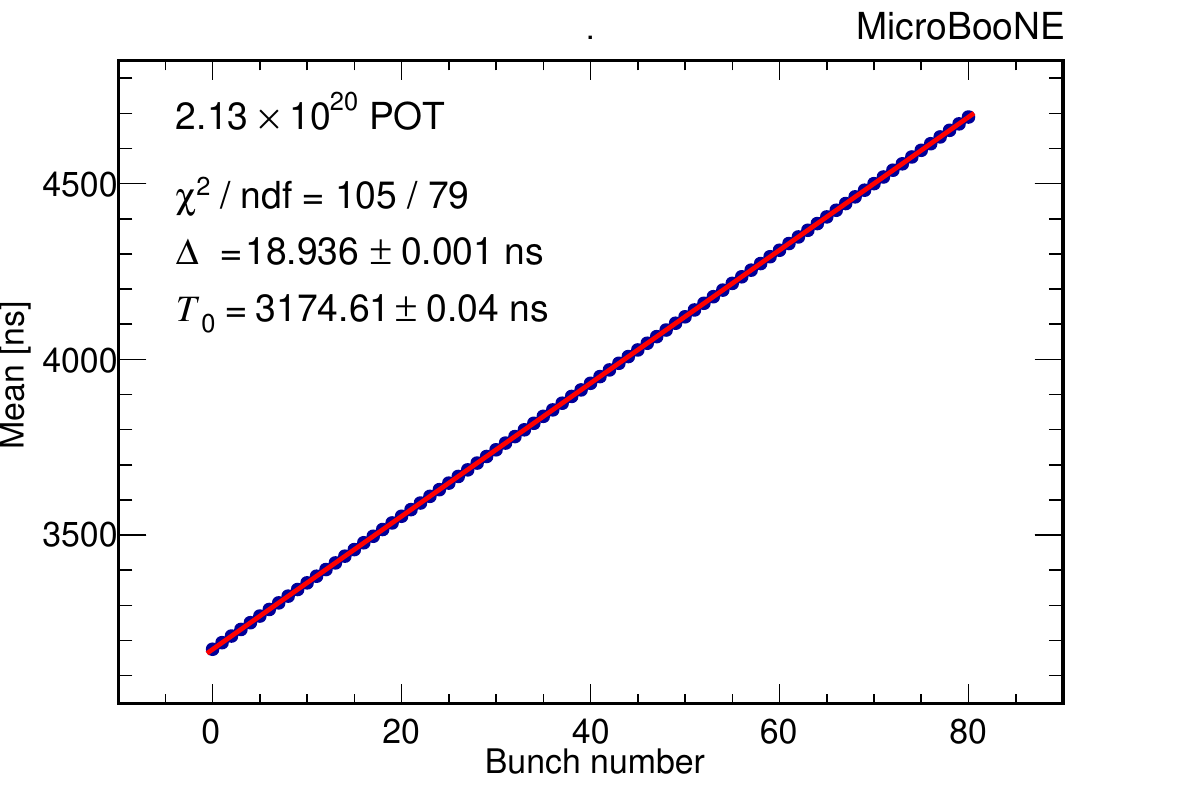}
\caption{\label{fig:linfit}For each of the 81 bunches observed in Fig.\ref{fig:med0Lb} (b) a Gaussian fit is performed to the bunch peak and the extracted mean values are used to obtain a linear fit as a function of the peak number. The gradient ($\Delta$) and the intercept ($T_0$) of the linear fit give respectively the bunch separation and the common constant offset due to the propagation time form the beam target to the TPC. The value found for the bunch separation is \unit[$\Delta = 18.936\pm 0.001$]{ns}.}
\end{figure}

 \emph{Neutrino arrival time reconstruction.} 
The neutrino arrival time, which is the neutrino time profile at the upstream detector wall, is reconstructed by removing the trigger jitter ($T$\textsubscript{RWM}), by subtracting from each PMT's measured time the neutrino ToF inside the TPC ($T_{\nu}$) and the daughter particle and photon propagation time ($T^*_{dp}+T^*_{sl}$), and by applying the empirical corrections ($T_{os}$ and $T_\mathrm{Emp}$). For each of these terms the spreads and the ranges of values are reported in Table~\ref{tab:spreadb}. It is important to note that a significant impact on improving the timing resolution comes from steps that make use of TPC reconstructed information emphasizing the importance of the analysis choice of leveraging both precise PMT timing and topological information from the TPC. Precise PMT timing is not alone sufficient to extract $\mathcal{O}$(\unit[1]{ns}) interaction timing resolution. The median of the obtained values across all PMTs with more than two detected photons is taken as the neutrino interaction time for the event. Figure~\ref{fig:med0Lb} shows the neutrino arrival timing before (a) and after (b) applying the neutrino interaction time reconstruction. The 81 bunches composing the beam pulse sub-structure are well visible after the reconstruction as seen in Fig.~\ref{fig:med0Lb} (b) and reproduce the \unit[52.81]{MHz} substructure of the RWM waveform of Fig.~\ref{fig:RWMstructure}. For each one of the 81 bunches a Gaussian fit is performed and the extracted mean values are used to obtain a linear fit as a function of the peak number, as shown in Fig.~\ref{fig:linfit}. The linear fit slope is used to measure the bunch separation ($\Delta$). The value found of \unit[$18.936\pm0.001$]{ns} matches the expectation from the accelerator frequency parameter~\cite{BNB2}. This work demonstrates for the first time $\mathcal{O}$(\unit[1]{ns}) timing resolution in neutrino interactions in a LArTPC using fully automated reconstruction methods which can be integrated in neutrino physics analyses. This analysis builds on past developments in the use of TPC and scintillation light information in LArTPCs, including previous work from ICARUS on neutrino time of flight measurements~\cite{ICARUS}.

\section{Results}
\label{sec:results}

\begin{figure}[H]
\centering
\includegraphics[width=0.48\textwidth]{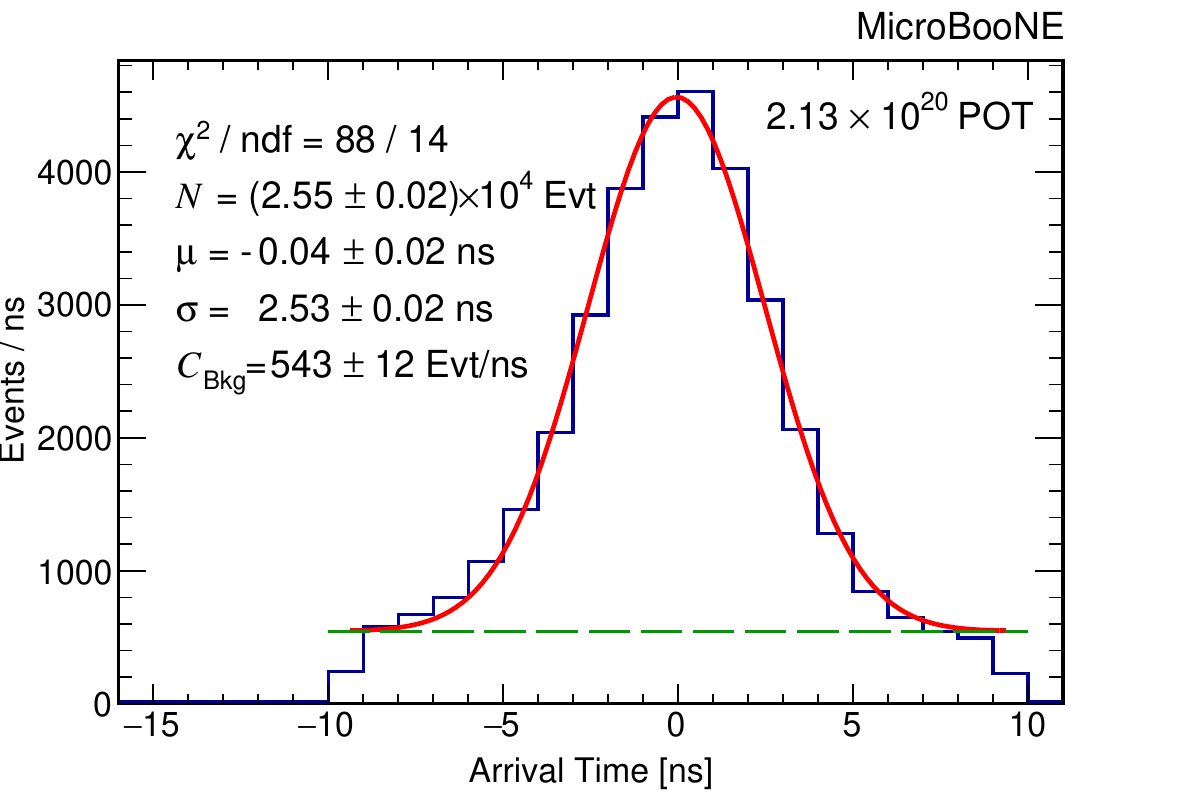}
\caption{\label{fig:med1a} Event timing distribution of the 81 beam bunches merged in a single peak after applying corrections. The green dashed line shows the constant term associated to the cosmic background uniform contribution.}
\end{figure}

\begin{figure}[H]
\centering
\includegraphics[width=0.48\textwidth]{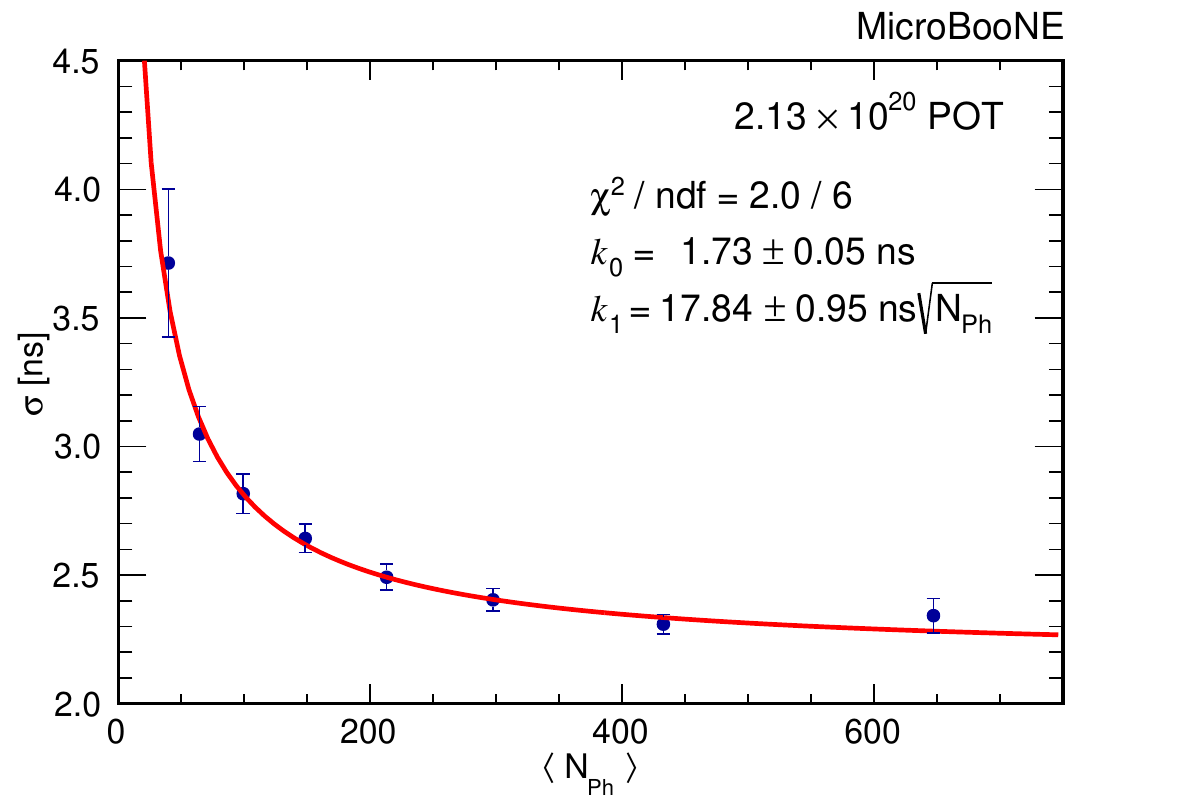}
\caption{\label{fig:med1b}Interaction timing resolution as a function of the total number of photons detected.}
\end{figure}

Once all the reconstruction steps are implemented and corrections applied, the neutrino candidate timing distribution, reported in Fig.~\ref{fig:med0Lb}~(b), is used to extract the detector timing resolution for neutrino interactions. The 81 bunches are merged in a single peak which is fit with the function:
\begin{equation}
\label{eq:3gfit}
\begin{split}
 \resizebox{.8\hsize}{!}{$ f(t)= C_\mathrm{Bkg}+ \frac{N}{  \sqrt{2 \pi \sigma^2} } \left\{ \exp{\left[ -\frac{1}{2} \left( \frac{ t- \mu -\Delta}{ \sigma } \right)^2 \right]} + \right. $} \\  \resizebox{.8\hsize}{!}{$ \left.
+ \exp{ \left[-\frac{1}{2} \left( \frac{ t- \mu  }{ \sigma } \right)^2 \right]} + \exp{\left[ -\frac{1}{2} \left( \frac{ t- \mu + \Delta }{ \sigma } \right)^2 \right]} \right\}$} 
\end{split}
\end{equation}
The fit function is composed of three Gaussians with identical width $\sigma$. The fit parameter $\sigma$ is used to extract the timing resolution, while the two Gaussians offset by the bunch separation $\Delta$ are introduced to account for events from neighboring peaks. Finally an overall constant term, $C_\mathrm{Bkg}$, accounts for a flat background from cosmic-ray events. Using this method the bunch width obtained is \unit[$\sigma = 2.53\pm0.02$]{ns}, from the fit shown in Fig.~\ref{fig:med1a}. Table~\ref{tab:spread2} shows the reduction of the bunch width after each reconstruction step is included. 
\noindent Subtracting the intrinsic proton beam bunch width \unit[$\langle \sigma_{BNB} \rangle \simeq$1.308]{ns} from the measured bunch width gives a value for the overall detector timing resolution of
\begin{equation}
R_{Tot}=\sqrt{\sigma^2 - \langle 
\sigma_{BNB} \rangle^2}=2.16\pm 0.02\, \text{ns}
\label{eq:resS1}
\end{equation}
Finally, a characterization of the timing resolution versus the total number of detected photons is performed. The parameter $\sigma$ is measured as a function of the total number of detected photons, as shown in Fig.~\ref{fig:med1b}. This distribution is fit using the function 
\begin{equation}
\sigma\left(\langle N_{Ph}\rangle \right)=\sqrt{  \langle \sigma_{BNB} \rangle^2+ k_0^2+\left(\frac{k_1}{\sqrt{\langle N_{Ph}\rangle}}  \right)^2  }\text{,}
\label{eq:res}
\end{equation}
where $k_0$ is a constant term, $k_1$ is associated to the statistical fluctuation in the number of detected photons ($\propto\sqrt{ N_{Ph}}$), and $\langle \sigma_{BNB} \rangle$ is the beam spread contribution to the resolution. The intrinsic detector timing resolution is associated with the constant term $k_0$ measured to be \unit[$1.73\pm 0.05$]{ns}.

\begin{table}[h]
\caption{\label{tab:spread2}This table shows the decrease of the bunches width ($\sigma$) after each reconstruction step is applied. Applying singularly $T$\textsubscript{RWM} or $T_{\nu}$ is not sufficient to separate the bunches and measure the width. The intrinsic \unit[1.308]{ns} beam spread is included in the $\sigma$ values reported in this table.}
\centering
\begin{tabular}{l r}
 \\
\hline \hline
Correction included  & $\sigma$ [ns] \\
\hline
$T$\textsubscript{RWM} or $T_{\nu}$                                                      & -                   \\
$T$\textsubscript{RWM} and $T_{\nu}$                                                     &  4.7  $\pm$ 0.2     \\
$T$\textsubscript{RWM}, $T_{\nu}$, ($T^*_{dp}+T^*_{sl}$)                                 &  3.08 $\pm$ 0.04     \\
$T$\textsubscript{RWM}, $T_{\nu}$, ($T^*_{dp}+T^*_{sl}$), $T_{os}$                       &  2.99 $\pm$ 0.04     \\
$T$\textsubscript{RWM}, $T_{\nu}$, ($T^*_{dp}+T^*_{sl}$), $T_{os}$, $T$$_\mathrm{Emp}$                 &  2.53  $\pm$ 0.02    \\
\hline \hline
\end{tabular}
\end{table}


\section{Application of $\mathcal{O}$(\unit[1]{ns}) timing in Physics analysis}
\label{sec:app}
The $\mathcal{O}$(\unit[1]{ns}) timing resolution achieved can significantly expand MicroBooNE's capability of studying neutrino interactions and searching for BSM physics. An improved neutrino selection efficiency can be obtained by adding the $\mathcal{O}$(\unit[1]{ns}) timing as a new tool for cosmic background rejection in surface LArTPCs orthogonal to existing techniques~\cite{bib:SBN,bib:FM1,bib:FM2,bib:ArCC}. Moreover, a $\mathcal{O}$(\unit[1]{ns}) timing resolution allows improvement in the performance of searches for heavy long-lived particles which will travel to the detector more slowly than the SM neutrinos. This method can in particular be applied to searches for heavy neutral leptons (HNLs), expanding the phase-space and sensitivity of HNL models being tested with current techniques~\cite{bib:A2,HNL1}. In this section we describe the potential that the precise timing has for improved cosmic background rejection and for searches for heavy long-lived particles such as HNLs.

\begin{figure}[h]
\centering
\includegraphics[width=0.45\textwidth]{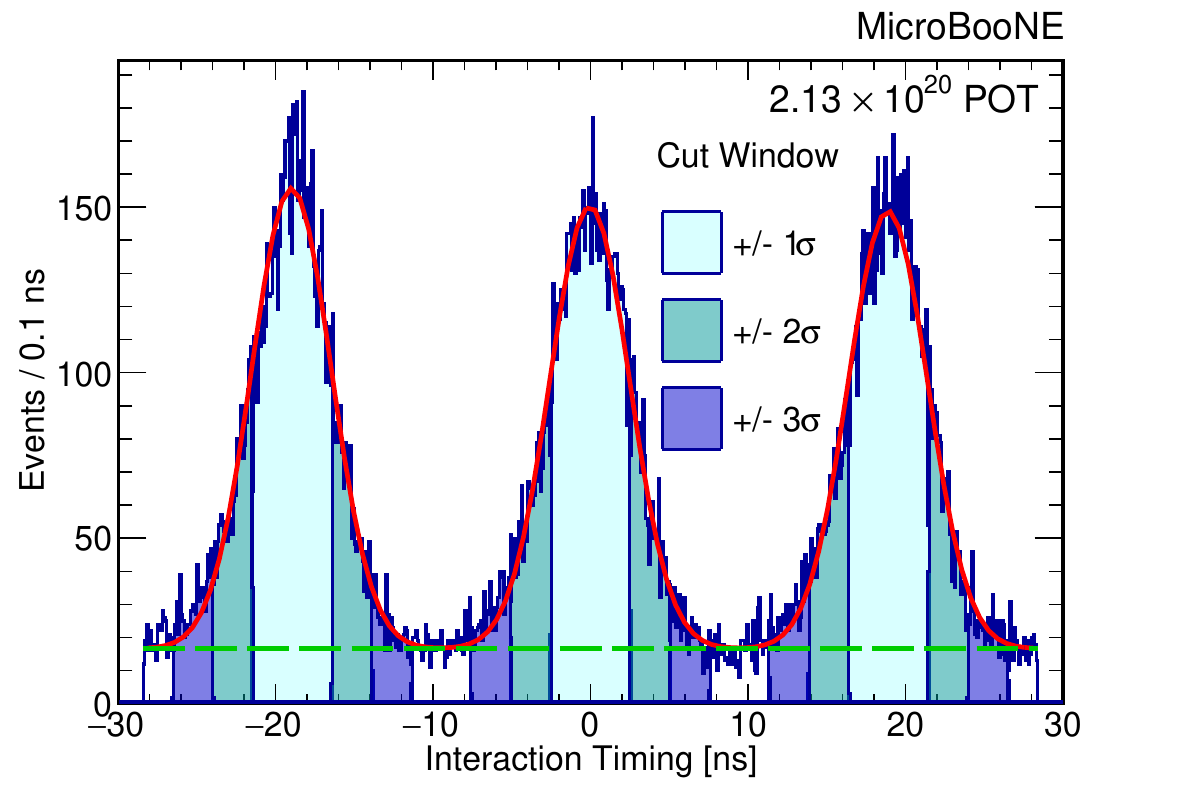}
\centering
\parbox{0.45\textwidth}{\centering(a) Event timing distribution with selection cuts around the peaks. The dotted green line shows the cosmic background fraction.}
\includegraphics[width=0.45\textwidth]{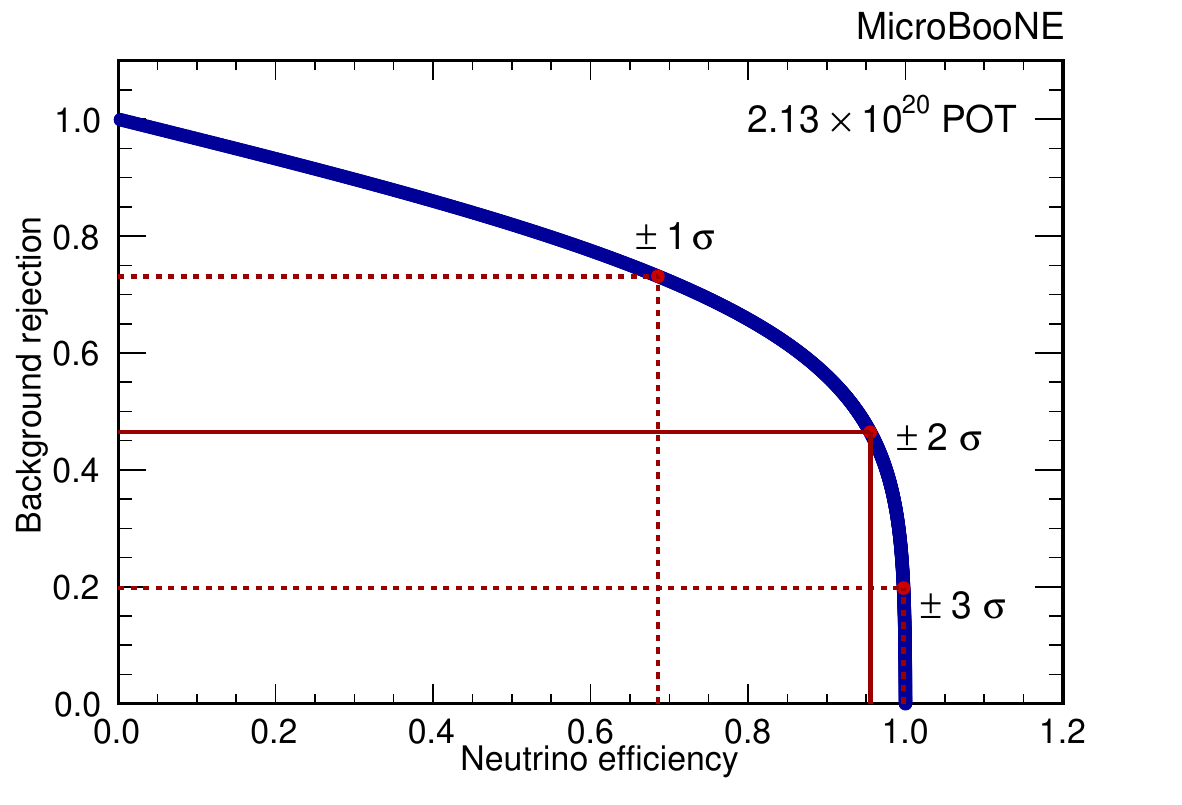}
\centering
\parbox{0.45\textwidth}{\centering(b) Neutrino efficiency versus background rejection.}
\caption{For the three cuts of  $\pm 3\sigma$, $\pm 2 \sigma$, $\pm \sigma$ around the peak the initial 27.1$\%$ of total background reduces to 21.7$\%$, 15.2$\%$, 10.6$\%$. Neutrino efficiency of 68.3$\%$, 95.5$\%$, 99.7$\%$ and background rejection of 73.3$\%$, 46.6$\%$, 19.8$\%$ are obtained for the respective cuts.}
\label{fig:bgr0}
\end{figure}

\begin{figure}[]
\centering
\includegraphics[width=0.48\textwidth]{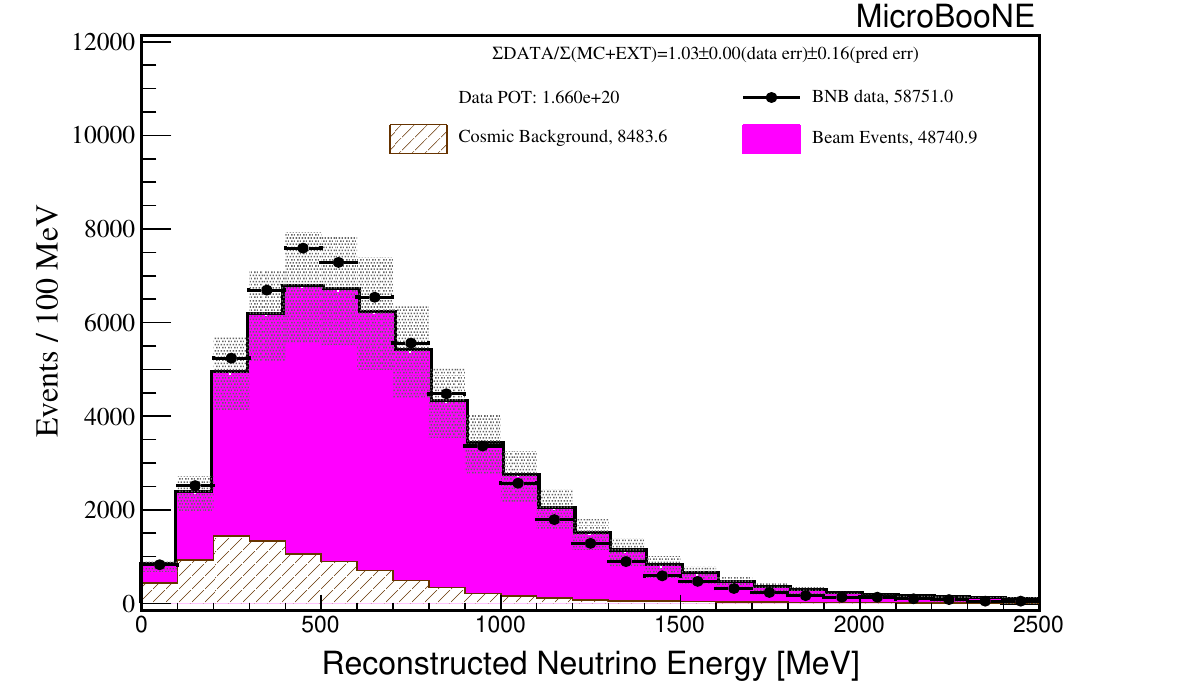}
\centering
\parbox{0.45\textwidth}{\centering(a) No additional cosmic removal cut around the interaction timing peak.}
\includegraphics[width=0.48\textwidth]{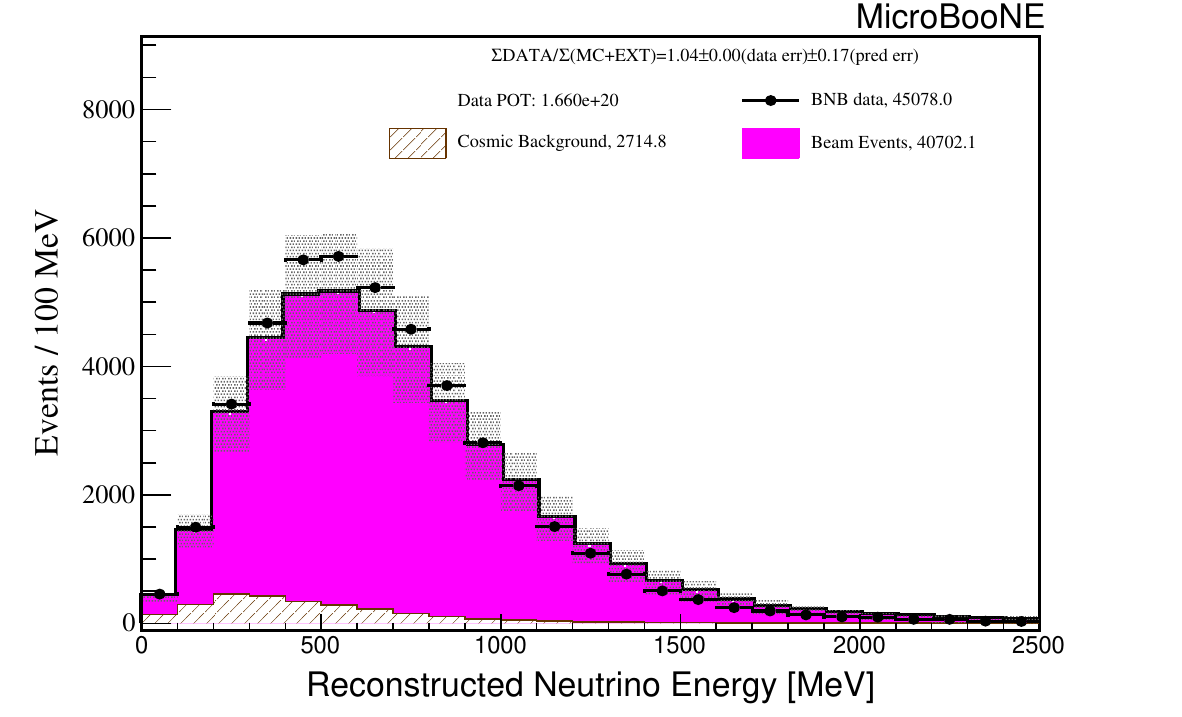}
\centering
\parbox{0.45\textwidth}{\centering(b) Cosmic removal cut of $\pm 2 \sigma$ around the interaction timing peak.}
\vspace{1em}
\caption{Reconstructed neutrino energy spectrum for events after Wire-Cell cosmic background rejection with (b) and without (a) an additional cosmic removal cut of $\pm 2 \sigma$ around the interaction timing peak.}
\label{fig:bgr3}
\end{figure}
\vspace{1em}

\par \emph{Cosmic ray background rejection.} As a surface-level LArTPC, cosmogenic backgrounds are a significant issue for MicroBooNE. Existing cosmic rejection techniques have achieved greater than 99.999$\%$ cosmic rejection while retaining greater than 80$\%$ of charge-current neutrino events~\cite{bib:FM1}. Nonetheless, these topology-driven techniques have significantly reduced performance for low-energy (less than about \unit[200]{MeV}) and neutral-current events. Additionally, even with greater than 99.999$\%$ cosmic rejection,  a cosmic contamination of 14.9$\%$ remains for a visible energy region greater than \unit[200]{MeV}, with closer to 40$\%$ contamination below \unit[100]{MeV}~\cite{bib:FM1}. Given this, cosmogenic backgrounds are often still the first or second largest background for MicroBooNE analyses~\cite{bib:PeLEE, bib:WCLEE, bib:last}, even when using the most up-to-date cosmic removal techniques~\cite{bib:FM1}. The reconstruction of the BNB bunch structure allows to exploit the timing of the neutrino interaction to reduce remaining cosmic-ray background. This is possible because cosmic-rays arrive uniformly in time while BNB neutrinos are in time with the proton pulse structure of Fig.~\ref{fig:RWMstructure}. Imposing a selection time window around the BNB bunches can be used to reduce the fraction of cosmic background events as shown in Fig.~\ref{fig:bgr0}~(a). Figure~\ref{fig:bgr0}~(b) shows the direct dependence of neutrino the selection efficiency versus background rejection. The neutrino selection efficiency is defined as the fraction of neutrino events surviving the cut applied to remove the background. As a benchmark, a cut at $\pm 2\sigma$ around the peak gives a $\nu_{\mu}$CC selection efficiency of 95.5$\%$ and a cosmic background rejection of 46.6$\%$ removing nearly half the cosmic-ray background with minimal efficiency loss. This method is complementary with respect to previously demonstrated cosmic rejection for LArTPCs which relies on charge-to-light matching~\cite{bib:FM1}. Figure~\ref{fig:bgr3} shows a demonstration of this method applied to the reconstructed energy spectrum for charged-current neutrino interactions from MicroBooNE.
The top panel shows current performance applying previous cosmic rejection techniques, while the bottom panel includes the neutrino interaction timing cosmic rejection developed in this work.

\vspace{1em}
\par
\begin{figure}[b]
\centering
\includegraphics[width=0.48\textwidth]{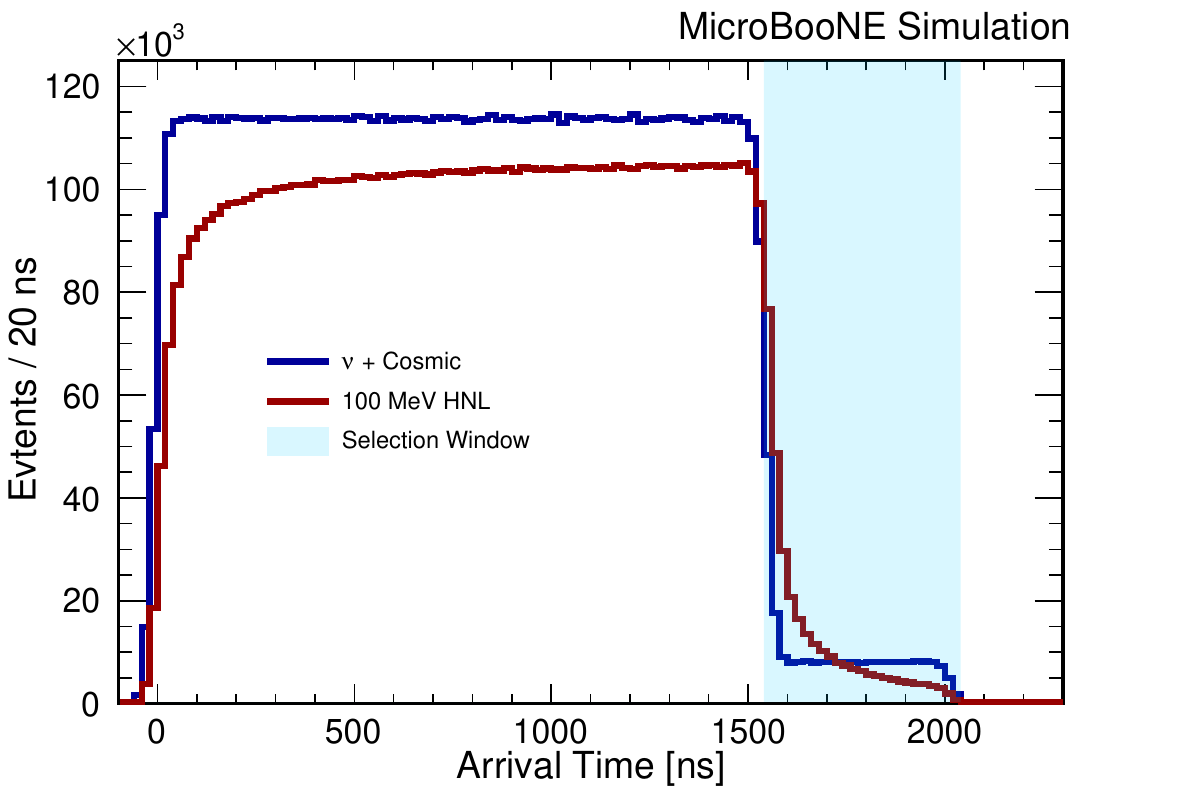}
\centering
\parbox{0.45\textwidth}{\centering(a) Timing information can be used to search for HNLs only in regions after the neutrino beam pulse when precise timing resolution is not available.}
\includegraphics[width=0.48\textwidth]{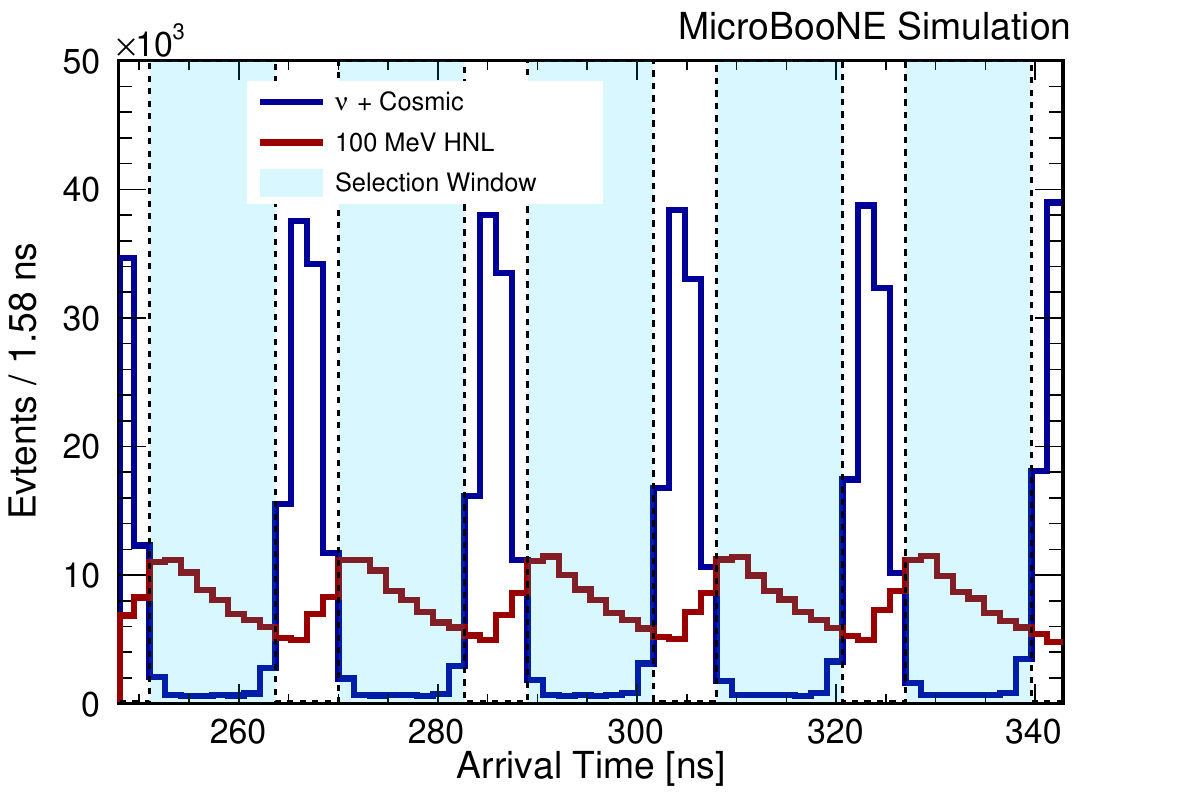}
\centering
\parbox{0.45\textwidth}{\centering(b) When the timing resolution can resolve the BNB substructure, each gap between the 81 bunches can be used to estimate the sensitivity to HNL.}
\vspace{1em}
\caption{Timing distribution for neutrinos and HNLs produced in the BNB. The ability to resolve the beam pulse substructure (b) offers significant improvement to the sensitivity in HNL searches compared to only the full \unit[1.6]{$\mu$s} pulse structure (a). This figure simulates an HNL with 100 MeV mass}
\label{fig:hnlbnb0}
\end{figure}

 \emph{Heavy Neutral Lepton Searches.} A set of models that can be tested with LArTPC neutrino experiments includes the production of HNLs through mixing with standard neutrinos~\cite{nup1,nup2,nup3,HNL1,bib:A2,bib:A1}. HNLs may be produced in the neutrino beam from the decay of kaons and pions, propagating to the MicroBooNE detector where they are assumed to decay to SM particles. The masses of these right-handed states can span many orders of magnitude, reaching the detector with a delay with respect to the nearly massless standard neutrinos~\cite{Ballett:2016opr}. This results in a distortion of the arrival time distribution when compared to the proton beam profile. To demonstrate the impact of ns timing resolution in HNL searches, the arrival time distributions of neutrinos and hypothetical HNLs at different masses and percentages are simulated. The BNB ns substructure measured in this analysis is used for both neutrino and HNLs assuming a \unit[1.5]{ns} timing resolution. HNLs are produced in the BNB with energies analogous to the neutrino flux. A 10$\%$ uniform cosmic background is included. Figure~\ref{fig:hnlbnb0} shows the arrival time distribution of standard neutrinos (blue line) compared to hypothetical  HNLs (red line) of \unit[100]{MeV} mass. When precise timing resolution is not available, timing information can be used to search for HNLs only in regions after the neutrino beam pulse, Fig.~\ref{fig:hnlbnb0} (a). When the timing resolution can resolve the BNB substructure, each gap between the 81 bunches can be used to estimate the sensitivity to HNL, Fig.~\ref{fig:hnlbnb0} (b). To quantitatively demonstrate the impact of timing resolution on HNL search sensitivity, a simulation study is carried out estimating signal and backgrounds for different HNL masses assuming only statistical uncertainties. The sensitivity in sigma is calculated using the Asimov sensitivity test given by
\begin{equation}
\label{eq:5st}
\resizebox{.9\hsize}{!}{$\sigma=\sqrt{2\left(s+b\right)\ln{\left(\frac{\left(s+b\right)\left(b+\sigma_b^2\right)}{b^2+\left(s+b\right)\sigma_b^2}    \right)}             -2\frac{b^2}{\sigma_b^2}     \ln{\left(1+    \frac{\sigma_b^2 s}{b\left(b+\sigma^2_b\right)}    \right)} }$}
\end{equation}

\begin{figure}[h]
\centering
\includegraphics[width=0.48\textwidth]{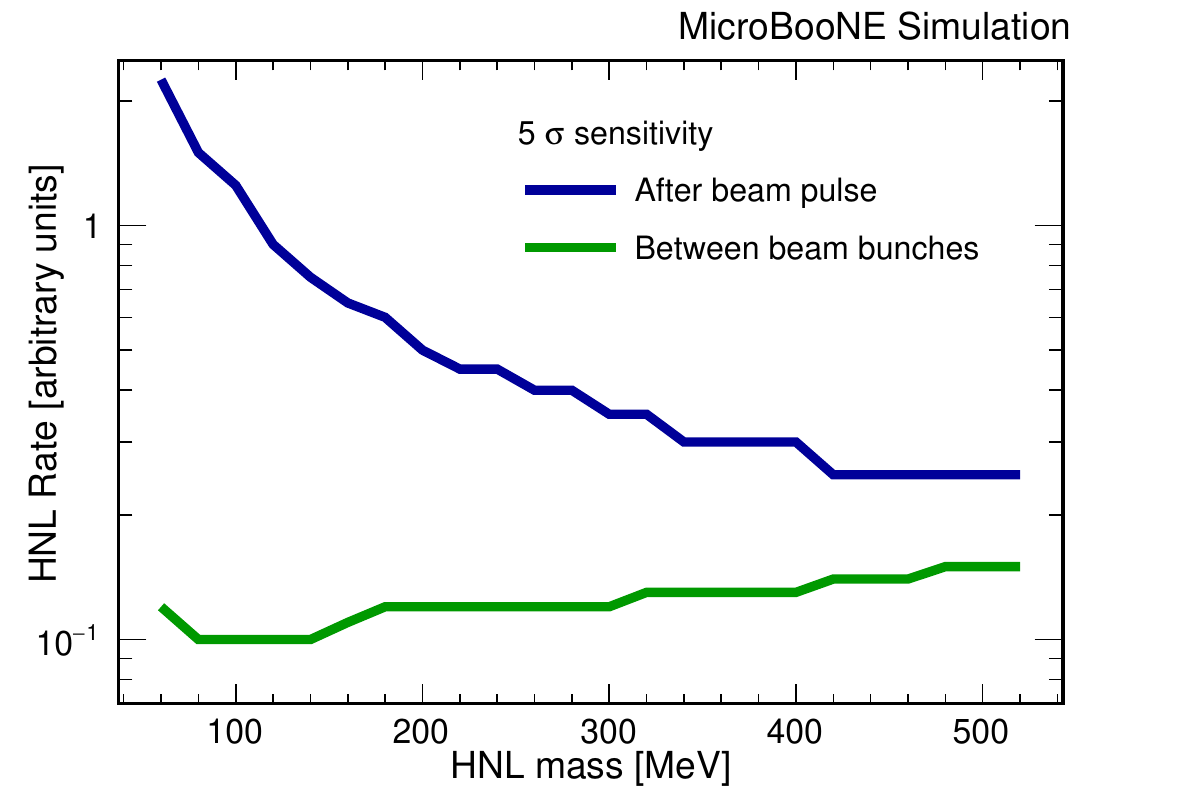}
\caption{\label{fig:5sl} Lines of 5 $\sigma$ sensitivity using only events after the beam pulse (blue line) compared to only events between beam bunches (green line), as function of the HNL mass. This study primarily focuses on the relative gain in sensitivity between the two methods as a proof of principle for future HNL searches.}
\end{figure}

\noindent where the signals (s) is the sum of the HNLs time distribution entries in a given windows and the backgrounds (b) is the sum of the BNB neutrino plus 10$\%$ of uniform cosmic background time distribution entries in the same windows, $\sigma_b$ is the standard deviation of the entries summed to obtain b. When using only events after the beam pulse the window used to estimate the sensitivity include time distributions entries from \unit[1540]{ns} to \unit[2040]{ns} (where the peak of the first neutrino bunch is centered at \unit[0]{ns}). When utilizing events between beam bunches, the included entries are in the gaps between neutrino bunches, in a window where the signal to background ratio is optimized to return the best sensitivity value. In this case the selection window size and position vary based on the mass, the bump shape and percentage of HNL simulated. This is done by first examining all regions with a non-zero HNL signal. Then a threshold for the minimum signal to background ratio is set that defines which bins shall be included in the window. This threshold is optimized to select windows between neutrino bunches that return the best sigma sensitivity as defined by the Asimov sensitivity test. Since these windows are defined based on an optimized threshold for signal to background ratio the threshold values and exact window sizes differ based on the HNL mass and percentage as these parameters change the exact arrival time of HNLs and overall signal values. Figure~\ref{fig:5sl} shows the 5 $\sigma$ sensitivity to HNL rate as function of the HNL mass, using only events after the beam pulse (blue line) compared to only events between beam bunches (green line). The beam bunches' resolution offers significant improvement overall, especially for lower masses. While a preliminary sensitivity study, this work demonstrates the significant physics impact that the methods presented in this paper will have in expanding the reach of searches for LLPs by up to an order of magnitude in poorly constrained regions of parameter space.

\vspace{1em}
The ability to resolve interaction timing with $\mathcal{O}$(\unit[1]{ns}) resolution introduces a new method to improve searches for long-lived particles (including HNLs) by rejecting neutrino backgrounds through the determination of the interaction time. This development will improve the sensitivity of and help expand the reach of BSM searches in the existing and upcoming accelerator-based neutrino physics program being carried out at Fermilab. In particular, the introduction of $\mathcal{O}$(\unit[1]{ns}) timing has the potential to allow model-independent searches for heavy long-lived particles for masses of 10s to 100s of MeV.



\section{Conclusions}

This work is the first demonstration of $\mathcal{O}$(1$\,$ns) timing resolution for reconstructing $\nu_{\mu}$CC interaction times in a LArTPC with the MicroBooNE experiment. This result is achieved through the implementation of novel analysis methods that measure and correct the ToF of neutrinos and their interaction products, as well as scintillation photons propagating through the detector volume. This makes use of both precise photon detection system timing resolution as well as detailed reconstructed TPC information to account for various delays in particle propagation through the detector. Moreover, the RWM signal has been used to improve the precision of the beam trigger. The analysis finds an intrinsic resolution in measuring the neutrino interaction time of $1.73 \pm 0.05$ ns. This result allows for the resolution of the pulse time structure of the BNB that, in turn, introduces a new powerful handle for physics measurements with LArTPC neutrino experiments. The method presented here can be applied to obtain $\mathcal{O}$(1$\,$ns) timing for any type of interaction occurring in the TPC. $\mathcal{O}$(1$\,$ns) timing resolution for neutrino interactions enables a new cosmic-rejection method to discriminate between neutrino interactions arriving in~$\sim$2~ns pulses in the BNB versus the continuous flux of cosmic-rays that constitute a significant background for surface-based LArTPC detectors. Furthermore, $\mathcal{O}$(1$\,$ns) timing accuracy can be leveraged for searches of BSM particles such as HNLs that have a longer ToF and reach the detector delayed with respect to neutrinos. The development of this new handle for studying BSM signatures will expand the sensitivity reach and parameter space that can be explored for searching for BSM signatures in LArTPC detectors operating in neutrino beams, both within the SBN program~\cite{bib:SBN} and in the DUNE near detector~\cite{bib:DUNEND,NDLAr}. 


\section*{Acknowledgements}

This document was prepared by the MicroBooNE collaboration using the resources of the Fermi National Accelerator Laboratory (Fermilab), a U.S. Department of Energy, Office of Science, HEP User Facility. Fermilab is managed by Fermi Research Alliance, LLC (FRA), acting under Contract No. DE-AC02-07CH11359.  MicroBooNE is supported by the following: the U.S. Department of Energy, Office of Science, Offices of High Energy Physics and Nuclear Physics; the U.S. National Science Foundation; the Swiss National Science Foundation; the Science and Technology Facilities Council (STFC), part of the United Kingdom Research and Innovation; the Royal Society (United Kingdom); and the UK Research and Innovation (UKRI) Future Leaders Fellowship. Additional support for the laser calibration system and cosmic ray tagger was provided by the Albert Einstein Center for Fundamental Physics, Bern, Switzerland. We also acknowledge the contributions of technical and scientific staff to the design, construction, and operation of the MicroBooNE detector as well as the contributions of past collaborators to the development of MicroBooNE analyses, without whom this work would not have been possible. For the purpose of open access, the authors have applied a Creative Commons Attribution (CC BY) public copyright license to any Author Accepted Manuscript version arising from this submission.

\bibliography{biblio}

\end{document}